\documentclass[aps,prx,superscriptaddress,twocolumn,eprint,amsmath,amssymb,longbibliography,nofootinbib]{revtex4-1}

\usepackage{amsmath,amsthm,amsfonts,amssymb,amscd}
\usepackage{fullpage}
\usepackage{lastpage}
\usepackage{graphicx}
\usepackage{hyperref}
\usepackage{wrapfig}
\usepackage{enumerate}
\usepackage{fancyhdr}
\usepackage{mathrsfs}
\usepackage{mathtools}
\usepackage{braket}
\usepackage{slashed}
\usepackage{bm}
\usepackage[dvipsnames]{xcolor}
\usepackage{float}
\usepackage[margin=0.55in]{geometry}
\usepackage[english]{babel}

\usepackage{subcaption}
\usepackage{comment}

\usepackage{tikz}
\usetikzlibrary{calc}
\usetikzlibrary{decorations.markings}

\usepackage[section]{placeins} 

\newcommand{\p}[2]{\ensuremath{\frac{\partial #1}{\partial #2}}} 

\newcommand{\beq}{\begin{equation}}
\newcommand{\eeq}{\end{equation}}

\newcommand{\mcm}[1]{\textcolor{black}{#1}}

\begin{document}

\title{Multi-defect Dynamics in Active Nematics}
\author{Farzan Vafa}
\email[Corresponding author:\ ]{fvafa@ucsb.edu}
\affiliation{Department of Physics, University of California Santa Barbara, Santa Barbara, CA 93106, USA}
\author{Mark J. Bowick}
\affiliation{Kavli Institute of Theoretical Physics, University of California Santa Barbara, Santa Barbara, CA 93106}
\author{M.~Cristina Marchetti}
\affiliation{Department of Physics, University of California Santa Barbara, Santa Barbara, CA 93106, USA}
\author{Boris I. Shraiman}
\email[Corresponding author:\ ]{shraiman@kitp.ucsb.edu}
\affiliation{Department of Physics, University of California Santa Barbara, Santa Barbara, CA 93106, USA}
\affiliation{Kavli Institute of Theoretical Physics, University of California Santa Barbara, Santa Barbara, CA 93106}

\date{\today}
\begin{abstract}

Recent  experiments and numerical studies have drawn attention to the  dynamics of active nematics. Two-dimensional active nematics flow spontaneously and exhibit spatiotemporal chaotic flows with proliferation of topological defects  in the nematic texture. It has been proposed that the dynamics of active nematics can be understood in terms of the dynamics of interacting defects, propelled by active stress. Previous work has derived effective equations of motion for individual defects as quasi-particles moving in the mean field generated by other defects, but an effective theory governing multi-defect dynamics has remained out of reach. In this paper, we examine the dynamics of 2D active nematics in the limit of strong order and overdamped compressible flow.  The  activity-induced defect dynamics is formulated as a perturbation of the manifold of quasi-static nematic textures explicitly parameterized by defect positions. This makes it possible to derive a set of coupled ordinary differential equations governing defect (and therefore texture) dynamics. Interestingly, because of the non-orthogonality of textures associated with individual defects, their motion is coupled through a position dependent ``collective mobility" matrix.   In addition to the familiar active self-propulsion of the $+1/2$ defect, we obtain new  collective effects of activity that can be interpreted in terms of non-central and non-reciprocal interactions between defects.
\end{abstract}

\maketitle


\section{Introduction}

Active nematics consist of collections of elongated units that consume energy to exert forces on their surroundings, while still tending to align, locally generating apolar liquid crystalline order~\cite{marchetti2013hydrodynamics,Simha2002,doostmohammadi2018active}. Nematic order has been reported in a number of two-dimensional realizations of active systems, including suspensions of cytoskeletal filaments and associated motor proteins~\cite{sanchez2012spontaneous,keber2014topology,kumar2018tunable}, epithelial monolayers~\cite{saw2017topological,kawaguchi2017topological,blanch2018turbulent}, and layers of vertically vibrated granular rods~\cite{narayan2007long}. When active forces are sufficiently high, active nematics exhibit spatio-temporally chaotic self-sustained flows that have been dubbed ``active turbulence'', where vortical flows are accompanied by the proliferation of topological defects in the nematic texture~\cite{giomi2013defect,thampi2013velocity,giomi2015geometry,doostmohammadi2017onset}. The relevant nematic defects are $\pm1/2$ disclinations, that are created and annihilated in opposite sign  pairs, with the resulting average defect density increasing with activity~\cite{doostmohammadi2018active}.

Previous theoretical work has made progress in formulating a description of ``turbulent'' active nematics by focusing on the dynamics of the topological defects as quasiparticles, with effective active interactions mediated by elastic distortions of the nematic texture and by active flows~\cite{giomi2013defect,keber2014topology,shankar2018defect,Shankar2019hydro}. This work and related ones~\cite{vromans2016orientational,tang2017orientation} have highlighted the importance of the anisotropy of  nematic disclinations, and in particular the role of the polarity  of the $+1/2$ defect by describing it as an effective active particle, with propulsive forces~\cite{narayan2007long,sanchez2012spontaneous,giomi2013defect,pismen2013dynamics} and aligning torques~\cite{shankar2018defect,Shankar2019hydro} determined by the active flows. 
Experiments and simulations of continuum active nematic hydrodynamics have suggested that active defects themselves exhibit emergent behavior and order in states with orientational order of defect polarity~\cite{decamp2015orientational,putzig2016instabilities,srivastava2016negative,Patelli2019,doostmohammadi2016stabilization,Pearce2020,Thijssen2020}. In spite of recent progress, the nature of this emergent behavior and its relevance to specific experimental situations remains largely not understood.

Much of the earlier work had focused on the limit where the distortions of the texture due to different defects can be treated as independent, an assumption that is at odds with the long-range nature of nematic elasticity.  An important open question is the role of multi-defect interactions in governing the defect dynamics. To that end, Ref.~\cite{cortese2018pair} 
obtained explicit solutions of the linearized equations determining quasi-static textures for neutral defect pairs and used them to describe defect pair-creation and annihilation. Generalization of the methods to multi-defect states is, however, cumbersome.   

In the present paper, we begin with the familiar hydrodynamic equations of a compressible active nematic film on a substrate and proceed by writing down the explicit quasistatic solution for a multi-defect nematic texture  fully parameterized by arbitrary position of $N$ defect cores. This forms a $2N$ dimensional ``inertial manifold" on which slow dynamics associated with defect motion unfolds. To derive general equations for the defect dynamics driven by activity, we consider a system deep in the nematic state and treat activity as a perturbation. Our analysis transforms the partial differential equations of active nematic hydrodynamics into a set of ordinary differential equations for the defect positions that fully incorporates multi-defect interactions and yields a number of new results. First we show that, even in the passive limit, the overdamped dynamics of defects as quasiparticles is governed by a non-diagonal mobility matrix that captures the fact that because of the overlap of the textures associated with different defects, motion of one defect effectively ``drags" the other defects, the source of the apparent non-locality being the long-range nature of elastic interactions in the nematic state. While the off-diagonal terms of the mobility matrix are small compared to the diagonal ones (that reduce to the well known defect friction~\cite{Denniston1996}), the off-diagonal terms fall off only logarithmically with interdefect distance. Activity renders the $+1/2$ defect self propelled along its axis, as shown earlier~\cite{narayan2007long,sanchez2012spontaneous,giomi2013defect,pismen2013dynamics}.  
It additionally generates new active forces among defects that are qualitatively different from the well-known  Coulomb interactions among defect charges. We also show that the forces on defects due to the active flow generated by all others are in general  non-central and non-reciprocal, and are controlled by multi-defect dynamics. Previous work by some of us~\cite{shankar2018defect} had obtained the effective dynamics of individual defects in the mean-field of other defects. In this approach, the orientation or polarization of the $+1/2$ defect was treated as an independent degree of freedom. Here, in contrast, we describe directly the dynamics of multi-defect textures without the need for the mean-field approximation. We show that in the deep nematic limit the polarization of a defect is not an independent degree of freedom, but it is directly determined by the position of all other defects. This provides a complete description of multi-defect dynamics, but yields defect-defect interactions that are intrinsically determined by the dynamics of all defects. Finally, our work makes explicit the nonreciprocal and non-central nature of the interaction between  defects,  a feature 
that has only recently begun to be appreciated~\cite{Maitra2020}.

The  paper is organized as follows. Sec.~\ref{sec:model} introduces the model and Sec.~\ref{sec:stationary} presents the class of quasi-stationary multi-defect solutions which we use to parameterize the dynamics of nematic textures. In Sec.~\ref{sec:method} we derive defect dynamics for the passive case, set up the perturbative scheme for the active case, and derive the defect dynamics equations including effects of active flow. In Sec.~\ref{sec:results}, we state our main results and their consequences for the multi-defect dynamics. Sec.~\ref{sec:discussion} presents the discussion of our results in a broader context. Most of the technical details are relegated to the Appendices A-E.

\section{The Model}
\label{sec:model}
We consider a two-dimensional nematic liquid crystal described by the the Landau-de Gennes (LdG) free energy ~\cite{chaikin2000principles}, ${\cal F}(\{{\bf Q}\})$,
\beq
{\cal F}(\{{\bf Q}\})={1\over 2} \int dxdy \left [ K  \ Tr ({ \nabla} {\bf Q})^2+g[1- 2 Tr ({\bf Q}^2)]^2 \right ]\;,
\eeq
with  $2D$ traceless tensor order parameter of the form
\beq
{Q}_{ab}=A[{\hat n}_a {\hat n}_b-{1 \over 2} \delta_{ab}]
\eeq
expressed in terms of the position dependent director field ${\mathbf{\hat n}}$.  The rigidity parameter, $K$, defines the energetic cost of spatial variation of ${\bf Q}$ (for simplicity we shall consider the single Frank constant approximation) and  $g$, with units of energy density, controls the strength of nematic order, via the coherence length $\xi=\sqrt{K/2g}$ controls spatial variations in the magnitude of the order parameter $A$. Below we assume to be deep in the nematic state ($g \rightarrow \infty$), where $\xi$ is smaller than all other relevant lengthscales. In this  limit $A\approx 1$ and  the magnitude of the order parameter $Tr ({\bf Q})^2 \approx 1/2$ {\it almost} everywhere, exceptions being the cores of nematic defects of size $\sim \xi$.

The dynamics of a nematic is controlled by the balance of relaxation towards the minimum of the LdG free energy and advection of the tensorial order parameter by flow $\mathbf{v}$, according to
\beq
\partial_t {Q}_{ab}+ {\bf v} \cdot \nabla {Q}_{ab}= { 1 \over 2} [ {\bf Q}, {\bm\omega}]_{ab}- \frac{D}{4K}  {\delta {\cal F} \over \delta {Q}_{ab} }\;,
\label{eq:Q}
\eeq
where the diffusivity $D$  governs  relaxation towards equilibrium and $  \omega_{ab} =\partial_a v_b-\partial_b v_a$ is the vorticity.
In an \emph{active} nematic, flow is
generated spontaneously by  local extensile (or contractile) activity described by the active stress tensor proportional to the order parameter $\sigma_{ab}={\tilde \alpha}  {Q}_{ab}$ \cite{marchetti2013hydrodynamics,Simha2002}. Here $\tilde \alpha$, with units of energy density, measures the strength of the  activity, with $\tilde\alpha>0$
($\tilde\alpha<0$) corresponding to contractile (extensile) activity.
Assuming that flow is generated solely by the texture-dependent active force balanced by substrate friction $\mu$, the flow velocity $\mathbf{v}$ is determined by the force balance equation, given by
\beq
\mu v_a={\tilde \alpha} \partial_b {Q}_{ab}\;.
\label{eq:v}
\eeq
In Eq.~\eqref{eq:Q} we have dropped the rate of strain alignment source term \cite{marchetti2013hydrodynamics}, because in $2D$ and in the friction dominated, overdamped limit described by Eq.~\eqref{eq:v}, its effect on dynamics can be represented by renormalizing the rigidity constant~\cite{srivastava2016negative,putzig2016instabilities}.

We will rescale time with $\tau=\ell^2/D$, where $\ell$ stands for the characteristic separation between topological defects that are generated by activity~\cite{sanchez2012spontaneous,giomi2013defect,thampi2013velocity}. We restrict ourselves here to the case where this length is much larger than the coherence length $\xi$, hence the density of defects is low. We rescale all length with $\ell$. Deep in the nematic regime where $\xi$ is very small compared to all other relevant length-scales, we define  $\epsilon=\xi /\ell \ll 1$. This small parameter will be helpful in organizing the perturbation theory. Finally, we define the dimensionless activity parameter  $\alpha= \tilde \alpha/4\mu D$.

Because our approach will be entirely based on complex analysis, we  introduce it from the outset by defining the complex positional coordinates $z=x+iy$ and ${\bar z}=x-iy$ and the complex order parameter \cite{deGennes1972,de1995physics}
\beq
Q=({Q}_{xx}-{Q}_{yy})+i2{Q}_{xy}=Ae^{i\theta}
\eeq
in terms of which the (dimensionless) LdG free energy has the form
\beq
{\cal F}(\{Q\})= 
\int dzd{\bar z} \left [4|\partial Q|^2+\epsilon^{-2}
(1- |Q|^2)^2 \right ]\;,
\eeq
where $\partial =\partial_z= { 1\over 2} [\partial_x-i\partial_y]$ (and ${\bar \partial} =\partial_{\bar z}= { 1\over 2} [\partial_x+i\partial_y]$).

In the complexified and rescaled form, $v=\alpha \partial Q$ and the dynamical equation is recast as 
\beq
\partial_t Q = 
\mathcal I(Q) = -{\delta {\cal F}(\{Q\}) \over \delta {\bar Q}}+\alpha \mathcal I_\alpha(Q)\;,
\label{eq:complexQ}
\eeq
where
\begin{align}
\mathcal I_a (Q)&=  -
(\partial Q\partial Q + {\bar \partial} {\bar Q}{\bar \partial} Q) + (\partial^2 Q- \bar\partial^2\bar Q) Q
\end{align}
represents the active drive obtained by eliminating the flow velocity in favor of $Q$. The 1st and the 2nd terms describe, respectively, the advection of the order parameter and its rotation by the vorticity.

\section{Stationary and quasi-stationary  textures  deep in the nematic state}
\label{sec:stationary}

Stationary textures in the limit of zero activity ($\alpha =0$) minimize the LdG free energy and hence solve~\cite{de1995physics,Pismen1999}
\beq
{\delta {\cal F} \over \delta {\bar Q}}= -4 {\bar \partial }\partial Q - 2\epsilon^{-2}(1- |Q|^2)Q =0\;,
\eeq
the imaginary and the real part of which read, respectively,
\beq
{\bar \partial }\partial \theta+{\bar \partial } \log A \ \partial  \theta+{\partial } \log A \ {\bar \partial}  \theta=0
\label{eq:theta}
\eeq
and 
\beq
A^2=1-{ 2\epsilon^2} \left [ (\partial \theta)^2 -A^{-1}{\bar \partial }\partial A\right ]\;.
\label{eq:A}
\eeq
Deep in the nematic state ($\epsilon \rightarrow 0$) and away from possible singularities, Eqs.~\eqref{eq:theta} and \eqref{eq:A} are approximately
\beq
{\bar \partial }\partial \theta=0+\mathcal O(\epsilon^2)
\eeq
and 
\beq
A^2=1-2 \epsilon^2 (\partial \theta)^2 +\mathcal O(\epsilon^{4})\;.
\eeq
Thus, to the leading order in $\epsilon$, interesting nematic textures correspond to non-trivial solutions of the Laplace equation ${\bar \partial }\partial \theta=0$. While there are no non-constant harmonic functions on the plane (that are bounded at infinity), such functions exists on a punctured plane and define the ``topological defect" solutions. The simplest solution has the form $ \theta =i \sigma \log ({{\bar z} \over { z}})$, with $\sigma = \pm {1 \over 2}$ corresponding to the well known $2D$ nematic charge of $\pm 1/2$ disclinations, corresponding to
\beq
Q=\psi (z, {\bar z})=A_c(|z|)\left (  {z \over {\bar z}} \right )^{\sigma}\;,
\eeq
with the amplitude $A_c(|z|)$ describing the defect core~\cite{Pismen1999}: $A_c(0)=0$ and $A_c(|z|) \approx 1$ for $|z| > a $, where $a\sim \mathcal O(\epsilon)$.

More generally, one can construct a multi-defect texture starting with a harmonic function on a plane punctured at points $z_i$ labeling the defect positions, as
\beq
 \theta (z, {\bar z}) =i \sum_i \sigma_i \log ({{\bar z} -{\bar z}_i\over { z-z_i }})+2\psi\;.
 \label{thetamany}
\eeq
where the constant $\psi$ defines the orientation of the director at infinity. 
Eq.~\eqref{thetamany} gives rise to the order parameter texture of the form
\beq
Q_0(z, {\bar z}|\{z_i\})={\cal A}(z, {\bar z}) \prod _i \left (  {z-z_i \over {\bar z}-{\bar z}_i} \right )^{\sigma_i} e^{i2\psi}\;,
\eeq
with ${\cal A}(z, {\bar z}) =1-2\epsilon^2 |\partial \theta |^2+\mathcal O(\epsilon^{4})$ away from $z_i$ and  ${\cal A}(z, {\bar z}) \approx A_c(|z-z_i|)$ for $|z-z_i| \sim \epsilon$. With the proviso of ``charge neutrality" $\sum_i \sigma_i=0$, this texture satisfies a fixed boundary condition $Q \rightarrow e^{i2\psi}$ as $|z| \rightarrow \infty$. 
We will assume that defects are separated by distances $\ell$ much larger than the core size $|z_i-z_j| \gg \epsilon $, in which case $|Q| \approx 1$ almost everywhere: the  ${\mathcal O} (\epsilon^2)$ correction to ${\cal A}$ outside defect cores can be viewed as a finite density ${\mathcal O} (\epsilon^2)$ correction. The multi-defect  texture  $Q_0 (z,{\bar z}|\{ z_i \})$ minimizes  ${\cal F}(Q)$ to order ${\mathcal O} (\epsilon^2)$ on the punctured plane with fixed $z_i$.
The free energy ${\cal F}_0={\cal F}(Q_0)$ can be written in terms of the  defect positions in the well-known form
\beq
{\cal F}_0 \approx 4\int dz d {\bar z}  \ |\partial \theta |^2  =-8 \pi \sum _{i \ne j} \sigma_i \sigma_j \log  {|z_j-z_i| \over a}+C
\label{eq:F-Coulomb}
\eeq
describing an effective Coulomb interaction between defect charges~\cite{chaikin2000principles}, where the constant $C$  stands for the  sum of the core energies of all defects. This of course means that, even in the absence of any ``activity", the defect cores will move to minimize the free energy ${\cal F}_0$.
Hence, $Q_0$ textures, while being extremal on a punctured plane, are only quasi-static. The manifold  of textures $Q_0 (z,{\bar z}|\{ z_i \})$ parameterized explicitly by $\{ z_i \}$ as independent ``collective" coordinates, defines the ``inertial manifold" on which the relatively slow dynamics due to defect interactions unfolds~\cite{Temam1990}.
Our analysis below provides the description of the slow dynamics within the inertial manifold and sets up a perturbative scheme for computing activity-dependent corrections to $Q_0$,  which, when projected on the inertial manifold, define the effective dynamics of $z_i$ with the inclusion of active forces. In using perturbation theory to define slow dynamics of defects, our work follows the well established paradigm of (extended) dynamical systems theory~\cite{Cross1993}  which was also used previously in the study of single defect dynamics~\cite{Denniston1996,Pismen1999,pismen2013dynamics,shankar2018defect}.

\begin{figure}[]
\centering
	\subcaptionbox{$+1/2$}
{\includegraphics[width=0.45\columnwidth]{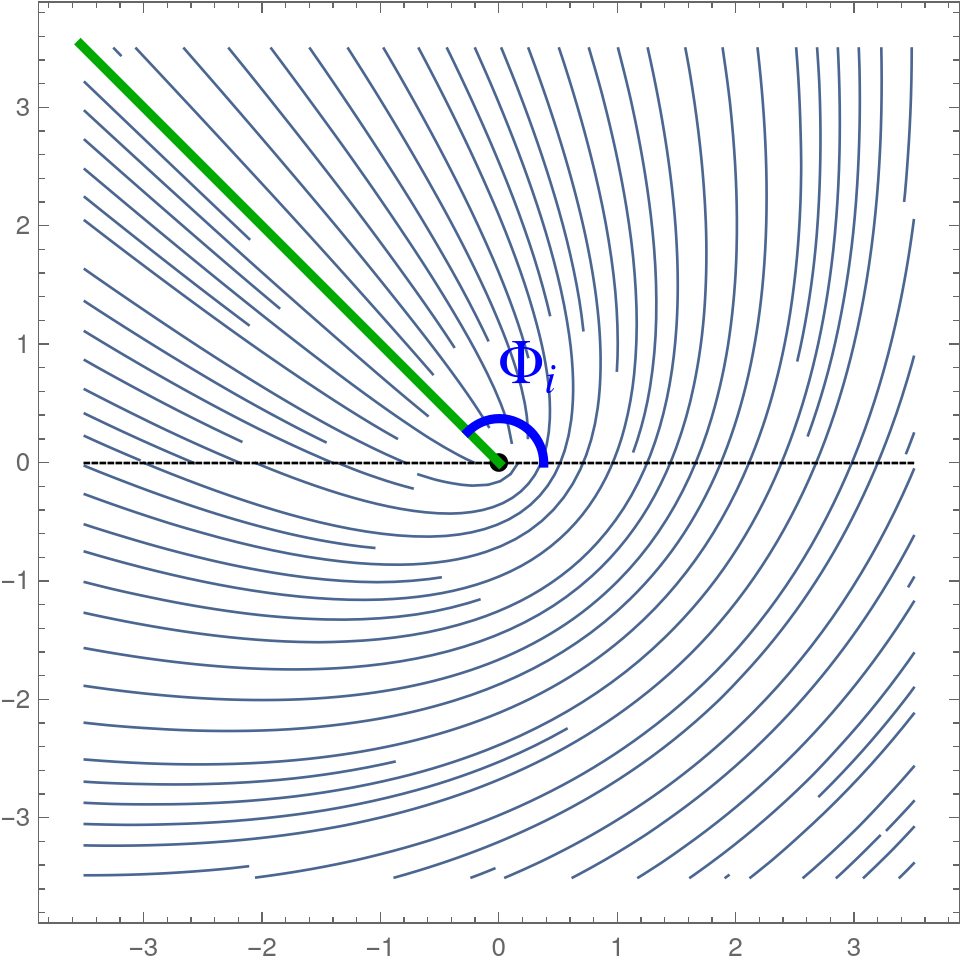}}
	\subcaptionbox{$-1/2$ defect}
{\includegraphics[width=0.45\columnwidth]{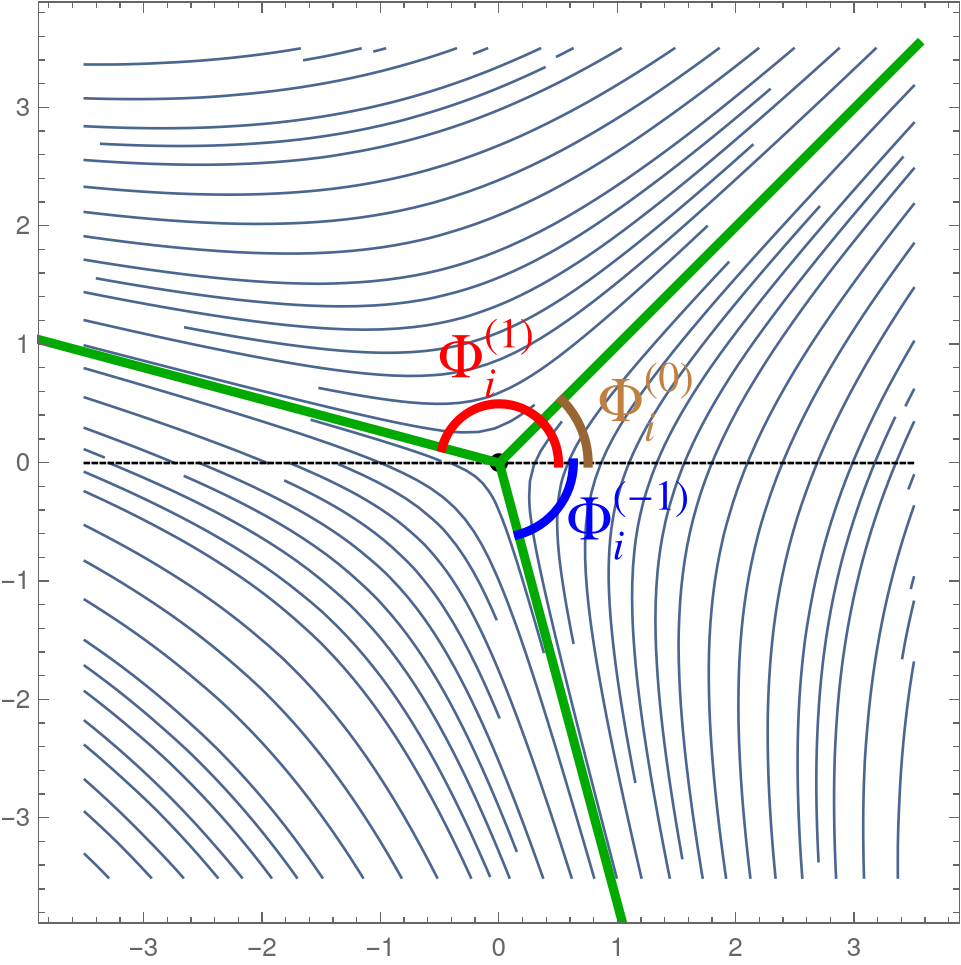}}
\caption{\mcm{Sketches of single defect textures showing the angles $\Phi_i$ for (a) a $+1/2$ defect where $\Phi_i = \phi_i = 3\pi/4$, and (b) a $-1/2$ defect where $\Phi_i^{(k)} = \phi_i/3 + 2\pi k/3$ for $k=0\pm1$.}}
\label{fig:pols}
\end{figure}

Before proceeding with the analysis, we note that in the vicinity of a defect, e.g. $z \approx z_i$, we can write
\beq
Q_0 (z, {\bar z}) \approx e^{i\phi_i}  \psi (z-z_i, {\bar z}-{\bar z}_i)\;.
\eeq
In other words, the texture reduces to the isolated defect form,
with a phase factor that depends on the positions of all defects (and the boundary condition at infinity)
\beq
e^{i\phi_i} =\prod _{j \ne i} \left (  {z_i-z_j \over {\bar z}_i-{\bar z}_j} \right )^{\sigma_j}  e^{i2\psi}  \;. \label{phase}
\eeq
This phase factor  will play an important role in controlling the active dynamics of defects.  It also readily interpreted in terms of the geometry of the director field close to the disclination, which exhibits  one ($+1/2$) or three ($-1/2$) separatrix lines emerging radially from the core, as shown in Fig.~\ref{fig:pols}. The separatrix is defined by the condition that the director points radially away from the core, which means that the polar angle of the director ${1 \over 2} \arg Q_0 |_{z \rightarrow z_i} =\varphi+\pi k$. Since the director angle in the vicinity of $z_i$ is ${1 \over 2} \arg Q_0 =\sigma_i \varphi +{1 \over 2} \phi_i $ where $\varphi =\arg (z - z_i)$, we can express the angle of the separatrix, $\Phi_i$, in terms of $\phi_i$ via $ \Phi_i =(\phi_i+2\pi k)/(2-2\sigma_i)$ which takes a unique value $\Phi_i =\phi_i$ for a plus disclination and three values
$\Phi_i^{(k)} =\phi_i/3+2\pi k/3$ (with $k=0,\pm 1$) for a minus disclination.

Last but not least, we note that for a global rotation, under which  $\varphi \rightarrow \varphi+\eta$ and $\psi \rightarrow \psi+\eta$, the complex order parameter transforms as $Q_0 \to Q_0 e^{i2\eta}$ , with the phase factor arising from the transformation of $\psi$. It follows that ``defect phases" $\phi_i$ transform in a way that depends on the associated charge $\phi_i \to \phi_i+2(1-\sigma_i )\eta$. In contrast, the phases $\Phi_i=\phi_i/(2-2\sigma_i)$ transform naturally (i.e., shift by the rotation angle $\eta$) under rotation.

\section{Derivation of the multi-defect dynamics equations}
\label{sec:method}

\subsection{Defect dynamics on the inertial manifold}
Before considering  active  motion of defects, let us examine the  relaxational dynamics on the inertial manifold governed by Eq.~\eqref{eq:complexQ} with $\alpha=0$. Since we have the approximate solution $Q_0 (z,{\bar z}|\{ z_i, {\bar z}_i \})$ which defines the free energy ${\cal F}_0(\{ z_i,{\bar z}_i \})\equiv{\cal F}(Q_0)$ as a function of collective coordinates via Eq.~\eqref{eq:F-Coulomb}, one expects ${\dot z}_i(t)={d \over dt} z_i$ to be determined by $\partial_i {\cal F}_0$. To simplify notation, we introduce a $2n$-dimensional (where $n$ is the number of defects) set of coordinates $\{ \zeta_k \}=\{ z_i,{\bar z}_i \}$ (with $\zeta_k$ for $k=1,\ldots,n$ standing in for $z_i$ and $k=n+1,\ldots,2n$ denoting ${\bar z}_i$)
\beq
{\dot \zeta}_k=-\sum_l m_{kl}{\bar \partial}_l {\cal F}_0\;.
\label{eq:zetak}
\eeq
Naively one may expect the mobility matrix to be diagonal, $m_{kl}=\delta_{kl}$, as would be the case if $z_i$ (and hence $\zeta_k$) were kinematically independent degrees of freedom, as often assumed~\cite{shankar2018defect}. This is not, however, the case, as we will see by deriving defect dynamics on the inertial manifold $Q_0$ directly from Eq.~\eqref{eq:complexQ}. Restricting Eq.~\eqref{eq:complexQ} to the $Q_0 (z,{\bar z}|\{ \zeta \})$ manifold gives
\begin{align}
\partial_t Q (z,{\bar z})&=  \sum_k 
{\dot \zeta}_k \partial_k Q_0(z,{\bar z}|\{ \zeta \})
=\nonumber\\
&=- \sum_a
 {\delta \zeta_k \over \delta {\bar Q}_0(z,{\bar z}) } {\delta {\cal F}_0 \over \delta \zeta_k}\;,
 \label{eq:Q0}
\end{align}
where we have formally used the chain rule of differentiation (and abbreviated $\partial_k =\partial_{\zeta_k}$) to evaluate $\delta {\cal F}_0 /\delta Q_0$. The subtlety is in defining ${\delta \zeta_k \over \delta {\bar Q}_0(z) }$. While ${ \delta {\bar Q}_0(z,\bar{z}) \over \delta \zeta_k } =\partial_k {\bar Q}_0(z,\bar{z})$ is unambiguous, the meaning of the former is not obvious. To define it, we recall that the variation $\delta Q(z,\bar{z})$ is taken in the $L^2$ norm so that one can think of  $\delta Q(z,\bar{z})$ as an $N$-dimensional vector (with $N \rightarrow \infty$) and of ${ \delta {\bar Q}_0(z,\bar{z}) \over \delta \zeta_a }$ as an $N \times 2n$ matrix. It is then natural to define ${\delta \zeta_k \over \delta {\bar Q}_0(z,\bar{z}) }$ by the pseudo inverse so that it obeys
\begin{align}
\delta_{kl}&=\int dz d{\bar z} \left [ {\delta \zeta_k\over \delta {\bar Q}_0(z,\bar{z})} { \delta {\bar Q}_0(z,\bar{z}) \over \delta \zeta_l } +{\delta \zeta_k\over \delta { Q}_0(z,\bar{z})} { \delta { Q}_0(z,\bar{z}) \over \delta \zeta_l } \right ] \nonumber\\
&=2\int dz d{\bar z}  {\delta \zeta_k\over \delta {\bar Q}_0(z,\bar{z})} { \delta {\bar Q}_0(z,\bar{z}) \over \delta \zeta_l }\;,
\end{align}
where the second equality derives from the fact that to  leading order $|Q_0|=1$ so that ${\bar Q}\delta Q =-Q\delta {\bar Q}$. This ``completeness condition"
is satisfied by
\beq
{\delta \zeta_k \over \delta {\bar Q}_0(z,\bar{z})} ={ \bar \partial}_l { Q}_0(z,\bar{z})  \mathcal M_{kl}^{-1}
\eeq
(where we have adopted the Einstein repeated index summation convention) with a Hermitian matrix
\beq
\mathcal M_{kl}= 2\int dz d{\bar z} \ \partial_k {\bar Q}_0 {\bar \partial}_l { Q}_0\;.
\label{metric}\eeq
The ``overlap" matrix $\mathcal M_{kl}$ can be thought of as the intrinsic metric of the inertial manifold parameterized by collective coordinates $\{ \zeta \}$. The explicit form of $\mathcal M_{kl}$ is computed in Appendix~\ref{app:M}.

Substituting into Eq.~\eqref{eq:Q0} we arrive at Eq.~\eqref{eq:zetak}
and we identify 
\beq
m_{kl}={\cal M}_{kl}^{-1}
\label{eq:mobility}
\eeq
relating the mobility matrix $m_{kl}$ to the metric ${\cal M}_{kl}$ of the inertial manifold.

\subsection{Defect dynamics in the presence of active stress.}
To describe the nematic dynamics in the limit of weak activity and low defect density, we will assume that the order parameter texture $Q(z,{\bar z}, t)$ stays close to the quasi-static manifold $Q_0 (z, {\bar z}|\{ z_i (t)\})$ parameterized by the time-dependent defect positions,
\beq
Q(z,{\bar z}, t)=Q_0 (z, {\bar z}|\{ \zeta (t)\})+\epsilon^2 \delta Q(z,{\bar z}, t)
\label{eq:Qsplit}
\eeq
where the $\epsilon^2$ prefactor of $\delta Q$ provides correct scaling for the magnitude of the perturbation, as we shall see below. We rewrite the complex texture dynamics, Eq.~\eqref{eq:complexQ}, as
\begin{align}
{\dot \zeta}_k \partial_k Q_0 &+\epsilon^2\partial_t \delta Q
=-{\delta {\cal F} \over \delta \bar Q}|_{(Q_0+\epsilon^2\delta Q)}
+\alpha \mathcal I_{\alpha}(Q_0+\epsilon^2\delta Q)\;.
\label{eq:dQ}
\end{align}
This serves as a starting point for our perturbation theory. Separating out variations within the inertial manifold described by the dynamics of  $\{  \zeta(t) \}$ from variations   described by $\delta Q$ on the punctured plane, we have
 \beq {\delta {\cal F}(\{Q\}) \over \delta \bar Q}={  \partial}_k { Q}_0 \mathcal M_{kl}^{-1} {\delta {\cal F}_0 \over \delta {\bar \zeta}_l} +\left [{\delta {\cal F}(\{Q\}) \over \delta \bar Q}\right ]_{\zeta}\;.
 \label{eq:dF}
 \eeq
Substituting Eq.~\eqref{eq:dF} into Eq.~\eqref{eq:dQ} and keeping terms to the linear order in activity and in $\epsilon^2$ leads to the linearized equation for $\delta Q$ in the form
\begin{align}
{\cal L}\delta Q+{\cal L}' & \delta {\bar  Q}=
{\cal V}_k \partial_k Q_0  - \alpha  \mathcal I_\alpha(Q_0) +{\cal O}(\epsilon^2 \alpha, \epsilon^2, \alpha^2) \;,
\label{eq:linear}
\end{align}
with
\beq
{\cal V}_k={\dot \zeta}_k +  {\cal M}_{kl}^{-1} {\bar \partial}_l {\cal F}_0 
\label{eq:Nuk}
\eeq
representing the residual motion along the inertial manifold. We have also introduced the linear operators
\begin{align}
&{\cal L} =2(1-2 |Q_0|^2)+4\epsilon^{2}\partial {\bar \partial}
\;, \\
&{\cal L}'= - 2 Q_0^2\;,
\end{align}
defined (on the punctured plane) by the linearization of ${\delta {\cal F} \over \delta Q(z) }(Q_0+\epsilon^2 \delta Q)$. We note that $\epsilon^2 \partial_t \delta Q$ does not appear in Eq. (29) to leading order in perturbation theory.
In the limit of $\alpha \to 0$, $\delta Q \to 0+{\cal O}(\epsilon^2)$ and ${\cal V}_k=0+{\cal O}(\epsilon^2)$ which corresponds to the passive dynamics described by Eq.~\eqref{eq:zetak} with $m_{kl}={\cal M}_{kl}^{-1}$. 

It is useful to reorganize Eq.~\eqref{eq:linear} and its complex conjugate into a matrix form,
\begin{align}
\left [
\begin{matrix}
&{\cal L} & {\cal L}'  \ \\
&{\bar {\cal L}}' & {\cal L} \ \\
\end{matrix}
\right ]
\left [
\begin{matrix}
&\delta Q \ \\
&\delta {\bar  Q} \ \\
\end{matrix}
\right ]=&
{\cal V}_k \left [
\begin{matrix}
& \partial_k  Q_0 \ \\
&\partial_k {\bar Q}_0 \ \\
\end{matrix}
\right ]
- \alpha \left [
\begin{matrix}
\mathcal I_{\alpha}\  \\
\mathcal {\bar I}_{\alpha}\  \\
\end{matrix}
\right ]\;.
\label{eq:matrix}
\end{align}

Crucially for our analysis, the linear operator on the left hand side of Eq.~\eqref{eq:matrix} has zero modes corresponding to infinitesimal changes of defect positions that enter $Q_0$ as free  $\zeta_k$ parameters, or
\begin{align}
\left [
\begin{matrix}
&{\cal L} & {\cal L}'  \ \\
&{\bar {\cal L}}' & {\cal L} \ \\
\end{matrix}
\right ]
\left [
\begin{matrix}
&\partial_k Q_0 \ \\
&\partial_k {\bar  Q}_0 \ \\
\end{matrix}
\right ]=0\;,
\label{eq:Lzero}
\end{align}
as can be seen explicitly by differentiating  ${\delta \mathcal F \over \delta Q }(Q_0)=0$ (which holds within our approximation on the punctured plane) with respect to $\zeta_k$. This means that $\delta Q$ corrections must be orthogonal to the inertial manifold $Q_0(z,\bar z |\{\zeta \})$, something that was already anticipated in representing the dynamics of $Q$ in the form given in Eq.~\eqref{eq:Qsplit}.

Hence the linear system of equations~\eqref{eq:matrix} for $\delta Q$ can only be solved if the inhomogeneous term is orthogonal to the null space of the linear operator, the so called ``Fredholm alternative" condition~\cite{Fredholm1903}. This solvability condition for Eq.~\eqref{eq:matrix} has the form
\beq  
\mathcal M_{kl} {\cal V}_l = \alpha  \int d^2z[\bar \partial_k \bar Q_0 {\mathcal I}_{\alpha} + \bar \partial_k Q_0 \bar {\mathcal I}_{\alpha}]\;,
\label{ODEApproximation}
\eeq
with the same overlap matrix $\mathcal M_{kl} $ as appeared in the previous section and (see Appendix~\ref{app:M}) 
\beq
 \int dz d{\bar z} \ {\bar \partial}_k {\bar Q}_0 { \partial}_l { Q}_0 =
 \int dz d{\bar z} \ { \partial}_k {\bar Q}_0 {\bar \partial}_l { Q}_0 = {1 \over 2} \mathcal M_{kl}\;.
 \eeq
Substituting the definition of $\mathcal V_k$, we arrive at the system of ODEs governing the defect dynamics, given by
\beq
 \mathcal M_{kl} {\dot \zeta}_l = -  {\bar \partial}_k {\cal F}_0+\alpha \ {\cal U}_k\;,
 \label{defect dynamics}
 \eeq
 with
 \beq
{\cal U}_k= \int dz d\bar{z}[\bar \partial_k \bar Q_0 {\mathcal I}_{\alpha} + \bar \partial_k Q_0 {\bar {\mathcal I}}_{\alpha}]\;.
\label{active force}
\eeq

Thus, flows driven by active stresses cause $Q$ to deviate from its form on the inertial manifold, with the deviation, $\delta Q$, limited by relaxational restoring forces. The component of active forcing that projects onto the tangent space of the inertial manifold shows up in the equations governing the dynamics of the defect positions $z_i (t)$. In the next section we will describe this dynamics more explicitly.

 An alternative derivation of the defect dynamics (Eq.~\eqref{defect dynamics}) is given in Appendix~\ref{app:altMethod}. We note also that the equations of motion for $z_i(t)$  obtained above minimize the deviation of the dynamics on the inertial manifold $Q_0$ from that described by the  equation of motion, Eq.~\eqref{eq:complexQ}. 
 In other words, the same dynamics can be obtained by   minimizing
\begin{align}
E &= \int d^2dz d\bar{z} \left| \partial_t Q( z,{\bar z},t) - {d \over dt} Q_0( z,{\bar z}|\{ z_i (t) \})\right|^2 \nonumber\\
&\approx \int dz d\bar{z} \left| \mathcal I (Q_0) - \dot z_i\partial_i Q_0 - \dot{\bar z}_i \bar\partial_i Q_0\right|^2
\label{eq:E}
\end{align}
with respect to $\dot z_i$, where $\mathcal I$ is defined in Eq.~\eqref{eq:complexQ}.

\section{Dynamics of defects in active nematics}
\label{sec:results}

In the previous section, we derived multi-defect dynamics based on a perturbation theory in defect density and activity. In this section, we present the explicit form of the resulting equations of motion, using the results for $\mathcal M_{kl}$ from Appendix~\ref{app:M} and for $\mathcal U_k$ from Appendix~\ref{app:U_i},
and discuss the nature of the various  terms. For clarity, we return to the original notation where we denote with $z_i$ and $\bar{z}_i$ the defect coordinates.

\subsection{Mobility matrix}

The matrix ${\mathcal M}_{k\ell}$ on the left hand side of Eq.~\eqref{defect dynamics} is  the inverse mobility matrix ~\cite{Brady1988}  representing the correlation in the motion of defects due to the non-orthogonality of the associated textures of order parameter. 
A nonlocal mobility is known to occur for colloidal particles in flow due to hydrodynamic interactions~\cite{Brady1988}. 
Evaluating Eq.~\eqref{metric} in Appendix~\ref{app:M}, we find that the $2n\times 2n$ matrix $\mathcal M_{k\ell}$ can be decomposed into four $n \times n$ blocks, and that the two off-diagonal blocks are much smaller than the diagonal parts, and hence we ignore it in the following. The two diagonal parts, which we denote by the $n\times n$ matrix $\mathcal M_{ij}$, are  equal and given by
\begin{align}
\mathcal M_{ij} &\approx 4\pi \sigma_i\sigma_j \ln \frac{L}{r_{ij}}\;,
\end{align}
where 
\beq r_{ij} = \begin{cases}
	|z_i - z_j|\ & i\neq j\\
	a\approx 0.8 \epsilon \ & i=j
\end{cases}\eeq
and $L$ is system size.  Only the diagonal part of $\mathcal M_{ij}$ receives contributions from the defect cores.

\subsection{Interactions due to active flows}

We next  present the result of evaluating the active forcing term $\mathcal U_i$ defined in Eq.~\eqref{active force}. In Appendix~\ref{app:U_i}, we show that
\begin{align}
\alpha{\cal U}_i = &\quad \pi \alpha a^{-1} e^{i\phi_i} \delta_{2\sigma_i,1} \ +\sum_{j \ne i} f_{ij}\;,
\label{active_U}
\end{align}
where
\beq 
f_{ij} = -2\pi\alpha \frac{\sigma_i\sigma_j}{1 - \sigma_j} { \bar q_{ij} - (-1)^{\delta_{\sigma_i + \sigma_j,1}}q_{ij} \over {\bar z}_i -{\bar z}_j }
\eeq
and
\beq 
q_{ij} = e^{i \phi_i}\left(\frac{z_i - z_j}{\bar z_i - \bar z_j}\right)^{\sigma_i-1} =  e^{2i(1-\sigma_i)\Phi_i}\left(\frac{z_i - z_j}{\bar z_i - \bar z_j}\right)^{\sigma_i-1}\;,
\eeq
with $\Phi_i = \frac{\phi_i}{2(1-\sigma_i)}$ the defect phase defined in Fig.~\ref{fig:pol}.

The first term in Eq.~\eqref{active_U} is the well-known ``self-propulsion'' of the  $\sigma_i=+1/2$ defect that arises from the flows that the defect itself generates ~\cite{giomi2013defect,pismen2013dynamics}, with the phase factor $e^{i \phi_i}$ controlling the direction. The latter is therefore recognized as the polarization (unit) vector of the $+1/2$ defect (see for e.g. \cite{vromans2016orientational,tang2017orientation,shankar2018defect}).

The second term describes forces induced by interaction with other defects, represented by the sum of pairwise terms, that like 2D Coulomb forces are inversely proportional to the pair separation $|z_i-z_j|$.  

Unlike Coulomb forces, pairwise forces here are in general non-reciprocal and depend on the relative positions of all other defects through the phase factor $e^{i\phi_{i}}$, thus incorporating many-body effects. Note that $\bar q_{ij} - (-1)^{\delta_{\sigma_i + \sigma_j,1}}q_{ij}$ is real, corresponding to a central force,  only in the case of $\sigma_i=\sigma_j=1/2$. In all other cases, this factor is purely imaginary, corresponding to active forces that act normal to the line joining defect positions, thus resulting in a  torque acting on the pair, that depends on the orientation of the pair relative to other defects and the order parameter in the far field.

Finally, for completeness we provide an explicit form of the multi-defect dynamics equations including both passive and active forces. After eliminating common factor of $4\pi$ from both sides, these are given by
\begin{align}
&\sum_k \left ( \sigma_i\sigma_k \log \frac{L}{r_{ik}} \right ) \dot z_k = 2\sum_{j\neq i}  \frac{\sigma_i \sigma_j}{{\bar z}_i-{\bar z}_j} + {\alpha e^{i\phi_i}  \over 4a}  \delta_{2\sigma_i,1}
\nonumber\\
&\quad  - \frac{\alpha}{2} \sum_{j\neq i}  \frac{\sigma_i \sigma_j}{(1 - \sigma_j)} { \bar q_{ij} - (-1)^{\delta_{\sigma_i + \sigma_j,1}}q_{ij} \over {\bar z}_i -{\bar z}_j }\;.
\label{def_dyn}
\end{align}
The three terms on the right hand side are, in order, the Coulomb interaction, the active self-propulsion of the $+1/2$ defects, and the active interactions.  This description of defect dynamics has a number of new features  discussed below.

\subsection{Non-centrality and non-reciprocity in active interactions}

In contrast to Coulombic interaction between defects, interactions mediated by active flow cannot be described by additive pair potentials. 
Nevertheless, active force acting on a given defect is represented as a sum of of pairwise terms which one interprets as a force exerted by one defect on another, even though this force depends in a specific way (through tensor $q_{ij}$ appearing in Eq.~\eqref{def_dyn}) on the global texture and hence on position of all other defects. 
  
  The active pairwise force term in Eq.~\eqref{def_dyn} has a non-trivial form and we now examine it in greater detail.
For a plus/minus disclination pair, we find that the force exerted on defect $i$ by defect $j$ is:
\beq 
f_{ij}=-{i \pi \alpha \over 1-\sigma_j}{  \sin[2(1-\sigma_i)(\Phi_i-\theta_{ij})] \over {\bar z}_i-{\bar z}_j }\ \ \ \text{for} \ \ \sigma_i \ne \sigma_j \;,
\label{eq:pmpair}
\eeq
where  $\theta_{ij}$ is the angle of the line joining $z_j$ to $z_i$ relative to the $x$-axis and we have used $\phi_i=2(1-\sigma_i)\Phi_i \, (\mod 2\pi)$. Notice that here, and in the following expressions,  $\Phi_i - \theta_{ij}$ is the relative angle between the polarization and line connecting the defects. This force acts perpendicular to the line connecting the defects and is clearly non-reciprocal since $\sigma_i \ne \sigma_j$, hence $|f_{ij}|\ne |f_{ji}|$. As a result, the disclination pair will experience a net force acting on its center of mass, as well as a torque which tends to rotate the pair until the line joining the defect centers aligns with the far-field phase. 
This is particularly clear in a system with just a single neutral disclination pair. In this case,  $2(1-\sigma_i)\Phi_i=2\sigma_j\theta_{ij}+2\psi$. Hence  $\sin[2(1-\sigma_i)(\Phi_i-\theta_{ij})]=\sin(2\psi-2\theta_{ij})$. 
Assigning $\sigma_i=1/2$, $\sigma_j=-1/2$ (and defining $z_i-z_j=r_{ij}e^{i\theta_{ij}}$) we have
 \begin{align}
 f_{ij} &= -\frac{1}{3}~\frac{2 i \pi\alpha e^{i\theta_{ij}}}{r_{ij}}\sin2(\psi-\theta_{ij})\nonumber\\
 f_{ji} &= \frac{2 i \pi\alpha e^{i\theta_{ij}}}{ r_{ij}}\sin2(\psi-\theta_{ij})\;.
 \end{align}
 
Due to the $1/(1-\sigma_j)$ prefactor in Eq.~\eqref{eq:pmpair}, the force acting on the plus-disclination $(i)$ is 3 times smaller than the force acting on the minus-disclination $(j)$, resulting in a net force acting on the center of mass, perpendicular to the axis of the pair. There is also a torque $T_{ij}={2 \over 3}|f_{ji}|r_{ij}$ that rotates the pair so as  to align
$\theta_{ij}$ with  $\psi$ - the order parameter orientation in the far field. 
Physically, this dynamics is due to the entrainment of the defects to the active flows generated by the global texture, with rotational invariance broken by the nematic orientation in the far field.
\begin{figure*}[]
\centering
	\subcaptionbox{$+1/2$ and $+1/2$ defects}
	{\includegraphics[width=.32\textwidth]{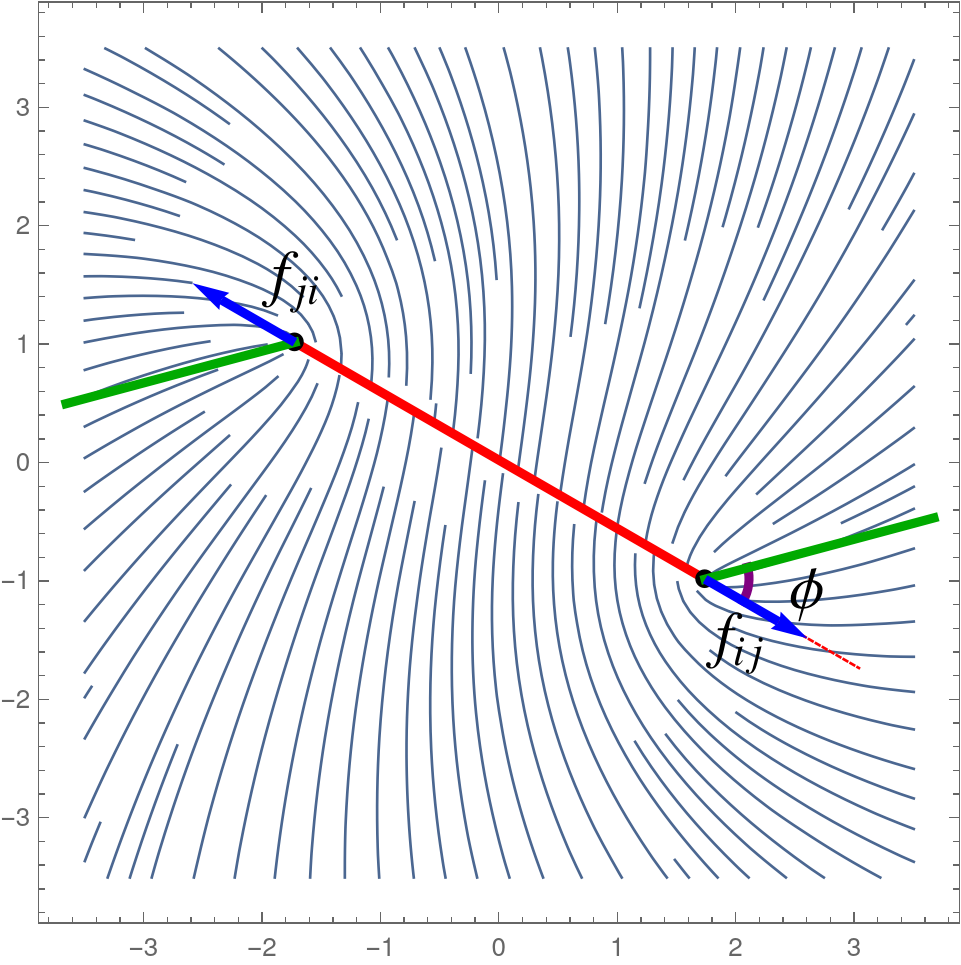}}
	\subcaptionbox{$-1/2$ and $+1/2$ defects}
	{\includegraphics[width=.32\textwidth]{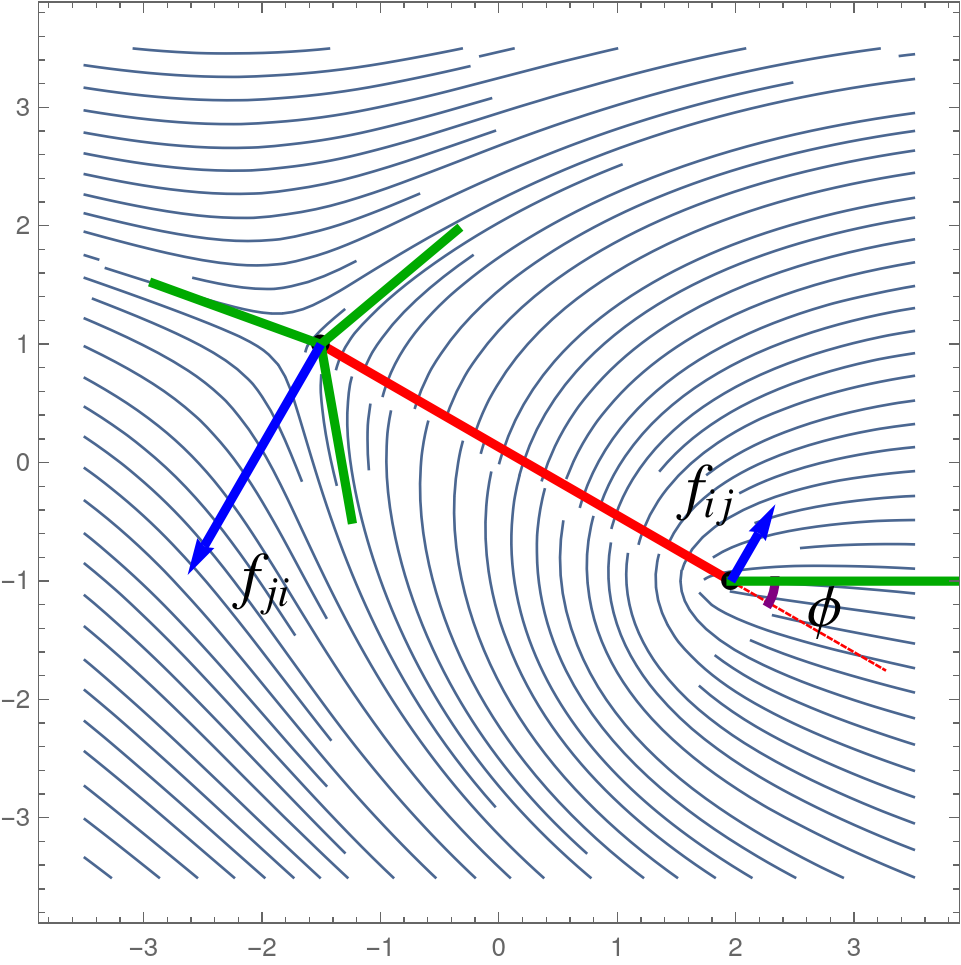}}
		\subcaptionbox{$-1/2$ and $-1/2$ defects}
	{\includegraphics[width=.32\textwidth]{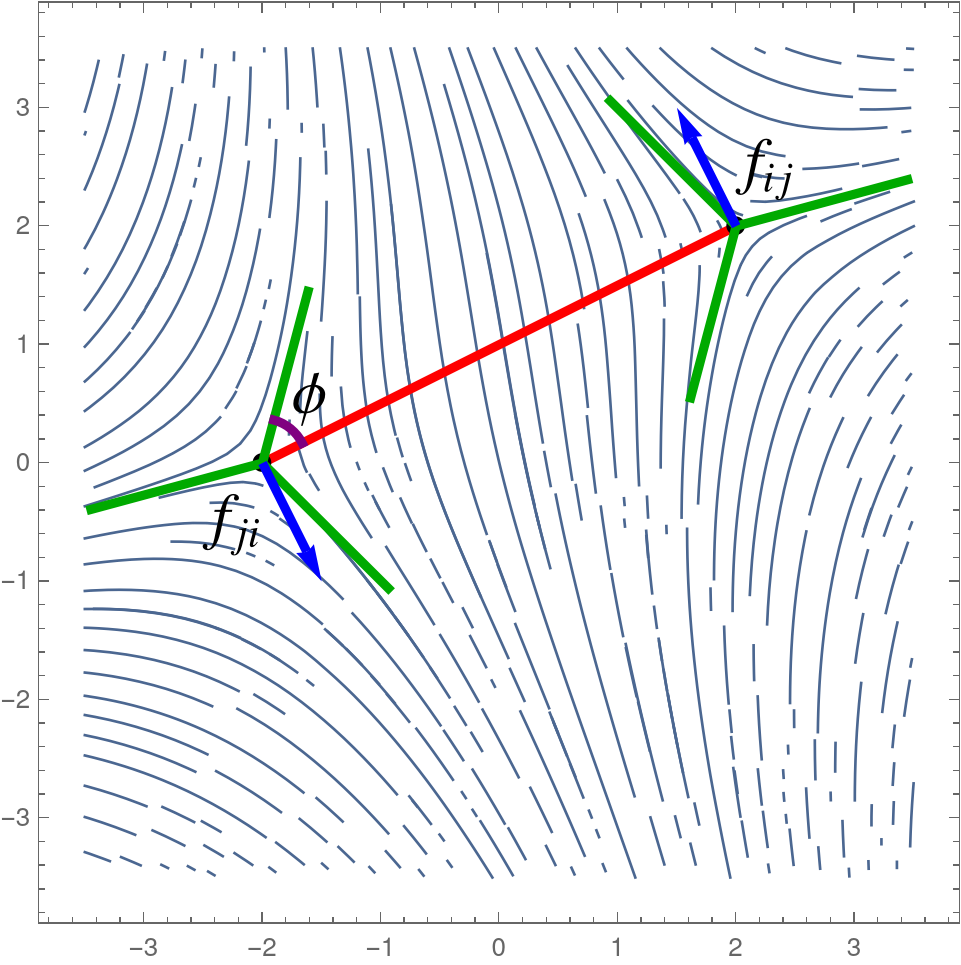}}
\caption{Sketches of the active forces between  defect pairs in an extensile system ($\alpha<0$). The blue arrows denote the forces, the red line joins the center of the two defects, and $\phi = \Phi_i - \theta_{ij}$ denotes the angle of the polarization relative to the line connecting the two defects. For $(+1/2, +1/2)$ pairs, the forces are radial, but for the $(-1/2,+1/2)$ and $(-1/2, -1/2)$ pairs, they are perpendicular to the line connecting the defects, generating rotations of the pair.}
\label{fig:pair-forces}
\end{figure*}

 The force on a plus disclination at $z_i$ arising from a second plus disclination at $z_j$  is
\beq 
f_{ij}={2\pi \alpha \cos(\Phi_i - \theta_{ij}) \over {\bar z}_i-{\bar z}_j }\ \ \ \text{for} \sigma_i=\sigma_j=1/2\;.
\label{eq:pppair}
\eeq
This force  acts along the line connecting the two defects. For a pair of plus-disclinations far away from all other defects, $\phi_i = \phi_j + \pi$, and so $f_{ij} + f_{ji} = 0$. Otherwise, $\phi_i \ne \phi_j + \pi$, and so $f_{ij}+f_{ji} \ne 0$. This lack of reciprocity is due to the gradient of the ``phase field" of the nematic texture.  The presence of other defects cannot be forgotten in this case because a pair of same sign defects alone does not satisfy the boundary conditions at infinity: at least two negative charge disclinations must be present to satisfy (topological) charge neutrality. We also note explicit dependence of $f_{ij}$ on the phase of $Q$ at infinity, $2\psi$, (which additively contributes to $\phi_i$). Rotating this phase would modulate the magnitude of $f_{ij}$ - an effect that is made plausible by noting that the same phase uniformly rotates the active stress tensor everywhere and hence rotates the direction of active flow relative to $z_i-z_j$.

Finally, for a pair of minus-disclinations  the force acting on $z_i$ is given by
\begin{align} f_{ij} 
= {i2 \alpha  \pi \over 3}{  \sin(3(\Phi_i-\theta_{ij})) \over {\bar z}_i-{\bar z}_j }\ \ \ \text{for} \ \ \sigma_i = \sigma_j =-{1 \over 2}\;.
\label{eq:mmpair}
\end{align}
Like the force in a neutral pair, this force also acts perpendicular to the line $z_i-z_j$, thus generating a ``2-body torque''. It is also non-reciprocal, thus yielding a net force acting on the pair.

The nonreciprocity and non-central character of the forces between  defect pairs arise because the texture, as described by the $Q$ tensor, is nonlinear in the director, hence in the defect phases. So, even if we write the texture as a linear superposition of individual defect phases, the flow generated by the texture is a nonlinear superposition of the flow generated by individual defects. As a result, the flow near one defect depends on the flow due to the other defects, and this results in the form of the forces and torques.

\subsection{Dynamics of $+1/2$ defect polarization}

\begin{figure}[]
\centering
\subcaptionbox{Two neighboring $+1/2$ defects}
{\includegraphics[width=.45\columnwidth]{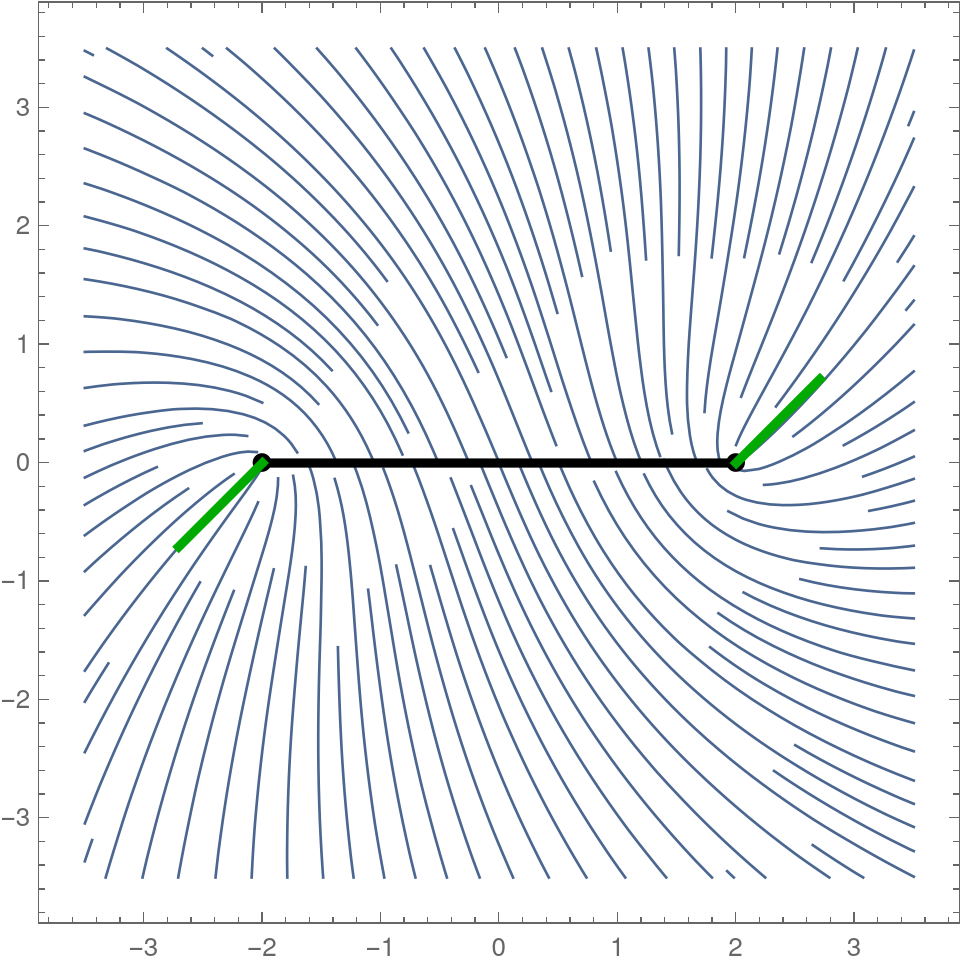}}
\subcaptionbox{Two $+1/2$ defects separated by a $-1/2$ defect}
{\includegraphics[width=.45\columnwidth]{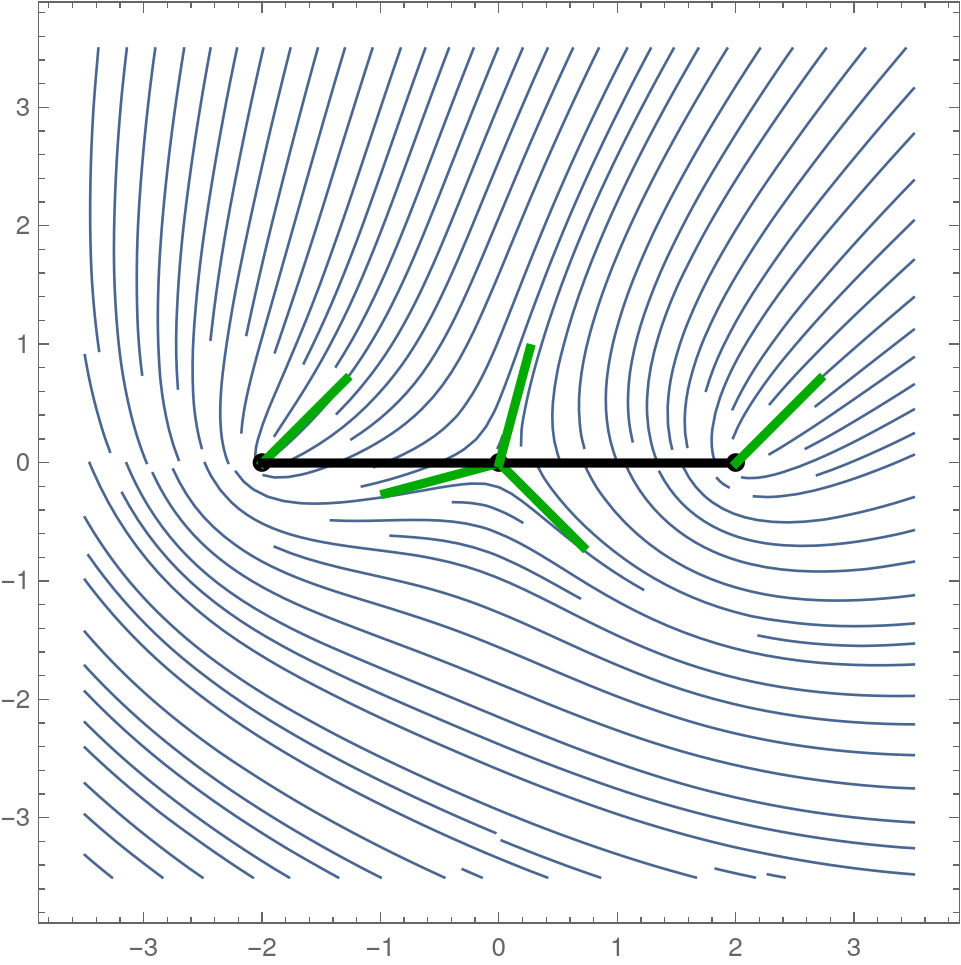}}
\caption{In (a), two neighboring $+1/2$ defects anti-align, and in (b), since the $+1/2$ defects are separated by a $-1/2$ defect, they align.}
\label{fig:pol}
\end{figure}

To illustrate our results and \mcm{make contact} with earlier work~\cite{keber2014topology,shankar2018defect,Shankar2019hydro}, we also construct an explicit equation for the dynamics of the polarization of a ``tagged'' $+1/2$ defect defined by the phase $\phi_i$ in the field of other defects. 
Differentiating Eq.~\eqref{phase} with respect to time, we obtain
\beq 
\frac{d\phi_i}{dt} = -i \sum_{j\neq i}\sigma_j \left(\frac{\dot z_i - \dot z_j}{z_i - z_j} - c.c\right)\;.
\label{phi_dyn}
\eeq
For simplicity, we evaluate this equation in the dilute limit, when interaction between defects (and the off-diagonal elements of the mobility matrix) can be neglected and defect motion is dominated by the active drift of plus-defects, with the result
\begin{align}
\frac{d\phi_i}{dt} &\approx -ic\alpha \sum_{j\neq i} \frac{\sigma_j e^{i\phi_i}}{z_i - z_j} +i c\alpha \sum_{j^+\neq i} \frac{ \sigma_{j_+} e^{i\phi_j}}{z_i - z_{j_+}} + c.c \nonumber\\
&= -2c\alpha \ {\cal E}_i\sin (\Theta_i-\phi_i) -2c\alpha \sum_{j_+ \ne i} {\sigma_{j_+}  \over r_{ij_+}}
\sin (\theta_{ij_+}-\phi_{j_+})\;,
\label{pol_dynamics}
\end{align}
where $c = [a\ln L/a]^{-1}$.  In the 1st term, we have defined
\beq
{\cal E}_ie^{i \Theta_i} = \sum_{j \ne i} {\sigma_j \over {\bar z}_i-{\bar z }_j}\;.
\label{eq:Efield}
\eeq
and in the 2nd term, the sum over ${j_+}$ runs only over plus-defects.
Up to a numerical factor, ${\cal E}_i$ is the ``electrostatic field"  at point $z_i$ due to other defects. The 1st term in Eq.~\eqref{pol_dynamics} describes the tendency of the plus-defect polarization in extensile (contractile) systems to align (anti-align) with the direction of net Coulomb force acting on it.
This term already appeared in the  equation for polarization dynamics derived in \cite{shankar2018defect} by examining the dynamics of a single defect in the mean field of other defects. It is purely kinematic in origin as it arises from  defect self-advection.  
The second term in Eq.~\eqref{pol_dynamics} is new.
The general structure of the polarization dynamics equation persists when defect dynamics ${\dot z}_i$ includes interaction terms of Eq.~\eqref{def_dyn}, which will cause the motion of minus-defects contributing to the second term in Eq.~\eqref{pol_dynamics}. Finally, we note in our formulation of the problem the equation for defect polarization dynamics is superfluous, as this polarization is defined kinematically by defect positions via  Eq.~\eqref{phase}. 

Explicit dependence of polarization on defect positions given by Eq.~\eqref{phase} provides some useful insights. For example, it is easy to see that two neighboring plus disclinations far removed from all other defects have their polarizations anti-align with each other (independent of the orientation of the pair axis). However, the presence of a minus-disclination (or plus-disclination) in between leads to the alignment of the polarizations of the two flanking plus-disclinations (see Fig.~\ref{fig:pol}).
These effects have been noted in earlier work on defect orientation~\cite{vromans2016orientational, tang2017orientation}, as well as in recent numerical and experimental studies~\cite{Pearce2020,Thijssen2020}.

\section{Discussion}
\label{sec:discussion}

We have presented a general formalism, based on perturbation theory in activity  and defect density, to derive a set of coupled ordinary differential equations governing multi-defect dynamics for a 2D active nematic deep in the nematic phase. Our analysis goes beyond  earlier work that obtained the dynamics of a single ``tagged'' defect in the mean-field of other defects~\cite{shankar2018defect}, to capture the coupled dynamics of a many-defect texture. This yields a number of new results and explicitly demonstrates the non-central and non-reciprocal nature of active stress-induced defect-defect interactions~\cite{Maitra2020}.

Central to our approach is the realization that deep in the nematic state, order parameter textures stay close to a 2N-dimensional ``inertial manifold" defined by the quasi-static multi-defect solution $Q_0$ parameterized by defect positions.
By explicitly describing multi-defect configurations, we obtain a closed formulation that describes defect dynamics entirely in terms of the defect positions. This avoids the need to treat polarization as an independent degree of freedom, as was done in
the earlier work by some of us~\cite{shankar2018defect}. 
Here the polarization of the $+1/2$ defect and the orientation of the $-1/2$ as defined by the ``defect phase" $\Phi_i$ are expressed explicitly in terms of defect positions through Eq.~\eqref{phase}.
Dynamics of the latter defines  the relatively slow  flow on the inertial manifold of textures by comparison to the rapid ($\sim {\cal O } (\epsilon^{-2})$) rate of relaxation of deviations away from the manifold. This separation of time-scales enables the perturbation theory about the $Q_0$ manifold. 

The separation of time scales that enables us to describe the dynamics of $Q$ by projecting it onto the inertial manifold holds as long as the mean separation between defects is large compared to the coherence length $\xi$  ($\epsilon^{2} \ll 1$). In our analysis here, the defect density was treated as given by  the initial condition. More realistically, nematic disclinations are subject to pairwise creation and annihilation. Extensive simulations of the continuum equations of active nematics have demonstrated that the state of spatio-temporal chaotic defect dynamics is characterized by a steady mean defect density $\ell^{-2} =\ell_{active}^{-2}\sim \alpha$~\cite{giomi2015geometry,hemingway2016correlation,doostmohammadi2018active}. For weak activity, this regime falls within the range of validity of our perturbation theory, even though we introduced it as an independent  double expansion in $\epsilon^{2}, \alpha \ll 1$.

One of the new results of our analysis is the recognition that defect velocities are directly coupled to each other through the inverse mobility matrix (on the left hand side of Eq.~\eqref{def_dyn}). This effect, while resembling  a ``collective drag",  is purely kinematic in origin as it arises from the non-orthogonality of order parameter deformations, $\partial_iQ_0$, associated with translational motion of individual defects. The overlap of these modes defines the metric of the inertial manifold parameterized by defect positions. Off-diagonal elements of the mobility matrix are only logarithmically smaller than the diagonal ones $ \sim \log (L/r_{ij}) / \log (L/a)$. 

Typical of 2D physics with soft modes, the mobility depends logarithmically on the size of the system $L$. Our formal way of dealing with the underlying infrared divergence has been to assume that defects are confined to a region much smaller than the size of the system $L$. It would however be straightforward to carry out the analysis in a finite disc, which can be done using the method of images as outlined in Appendix~\ref{app:finite}. Alternatively, our analysis can be carried out on the surface of a sphere or of a torus, where IR modes would be naturally cutoff by the radius. We note that the study of such geometries is directly relevant to experiments that have realized  active nematics on such curved surfaces~\cite{keber2014topology,ellis2018curvature}.

We note that non-diagonal mobility enters already in the purely relaxation dynamics of defects  driven by Coulomb interactions in the passive nematic.
Although the Coulomb interaction is reciprocal, so that the total force on the system of defects is exactly zero, the center of mass (COM) $\sum_i z_i$ can nevertheless move, provided that there are more than two defects (as can be seen directly from Eq.~\eqref{def_dyn}). This result may be counter intuitive, but it means simply that under relaxational dynamics, the defect system never translates rigidly as a whole and the motion of the COM is always accompanied by a change in defect configuration. By adding active forces with a suitable time dependence, it may be possible to take the system through a cycle that at the end restores relative position of the defects, but the possibility of a shift of the COM generated by such a ``stroke" would not be a surprise. 

Our analysis can be easily extended to include the interaction of defects with elastic deformations of the nematic order. To do that, one needs to replace the fixed phase $\psi$ in our definition of the $Q_0$ texture with an arbitrary harmonic function: $\Psi(z,t)+c.c$.
With this generalization, $Q_0$ would still minimize ${\cal F}$ on the punctured plane. $\Psi(z,t)$ allows to represent the effect of the external forcing acting and to compute finite size corrections as shown in Appendix~\ref{app:finite}.

Our work provides a unified framework for describing the dynamics of defects in active nematics and investigating the possibility of dynamical phases of defect order. Conflicting results have been reported in experiments and simulations with both polar and apolar (antiferromagnetic) order of the polar $+1/2$ defects reported by different authors ~\cite{decamp2015orientational,putzig2016instabilities,oza2016antipolar,srivastava2016negative,ellis2018curvature,shankar2018defect,Pearce2020,Thijssen2020}. While there have been suggestions that the type and range of defect order may be affected by the importance of density fluctuations and viscous dissipation, which are not included in the present calculation, the defects ODEs derived here could be used to settle some of these open questions. It would also be interesting to investigate the possibility of ordered lattices of defects accompanied by a regular array of flow vortices and resembling Abrikosov lattices in type-II superconductors that has been reported in simulations~\cite{doostmohammadi2016defect}. 

While this manuscript was being prepared, we learned of the preprint by Y-H. Zhang, M. Deserno and Z-C. Tu  posted on the arXiv~\cite{Zhang2020dynamics}. In this work, Zhang et al. derive - by a variational method similar to our Eq.~\eqref{eq:E} - active nematic defect dynamics equations for 4 plus disclinations on a sphere.
	
\begin{acknowledgments}
The authors thank Zvonimir Dogic, Eric Siggia, Suraj Shankar, Luiza Angheluta, Zhitao Chen, Supavit Pokawanvit, Zhihong You and the participants of the KITP Active20 program for stimulating discussions.  The work was supported by the NSF through grants PHY-1748958 (MJB), DMR-1609208 (MCM,FV) and PHY-0844989 (BIS). 
\end{acknowledgments}

\section*{Appendices}

\appendix

\section{Defect core structure}

Stationary textures in the limit of zero activity ($\alpha =0$) minimize LdG free energy and hence solve ~\cite{de1995physics,Pismen1999} 
\beq
{\delta {\cal F} \over \delta {\bar Q}}= -\nabla^2 Q - 2\epsilon^{-2}(1- |Q|^2)Q =0\;,
\eeq
the imaginary and the real part of which read, respectively,
\beq
\nabla^2 \theta+2\nabla \log A \cdot \nabla \theta=0
\eeq
and 
\beq
(1- A^2)={ \epsilon^2\over 2} \left [ A^{-1}\nabla^2 A-(\nabla \theta)^2 \right ]\;.
\eeq
We look for a solution for a single defect of charge $\sigma$ of the form
\beq Q = A(r)e^{2i\sigma\varphi}\;.\eeq
$A(r)$ would thus satisfy
\beq A''(r) + \frac{A'}{r} + \left(2\epsilon^{-2} - \frac{\sigma^2}{r^2} - 2\epsilon^{-2} A^2\right)A = 0\;.\label{A}\eeq
For example, for $\sigma = \pm 1/2$, $A(r)$ can be approximated as \cite{Pismen1999}
\beq A(r) = \tilde r \sqrt{\frac{.68 + .28\tilde r^2}{1 + .82 \tilde r^2 + .28 \tilde r^4}}\label{ASoln}\;,\eeq
where $\tilde r = r/\epsilon$. As $r\to0$, $A(r) \propto r$, and for $r \gg \epsilon$, $A(r) \simeq 1 - \frac{\epsilon^2}{4r^2}$. The defect core size $a$, which is the length scale over which $A$ goes from 0 to 1, is of the order $a \sim \epsilon$. As we will see later in the appendices, it is convenient for us to define the core size to be $a \approx 0.8\epsilon$.

\section{ Free energy of the multi-defect texture}
\label{app:LdG}

Here we compute the free energy in the passive case. For our ansatz $Q_0$, we have
	\begin{align}
	\mathcal F(Q_0) &= 4\int d^2z |\partial \theta|^2 = 4 \int d^2z \sum_{ij} \frac{\sigma_i\sigma_j}{(\bar z - \bar z_i)(z - z_j)} \nonumber \\
	&= \mcm{-8\pi \sum_{i\not=j} \sigma_i\sigma_j \ln \frac{|z_i - z_j|}{L}+C}\;,
	\end{align}
	\mcm{where $C$ is  a constant that accounts for the sum of the core energy of all defects}~\cite{chaikin2000principles}.
	So in the passive case we can view the free energy $\mathcal F$ as a function of defect positions $z_i$.
	
\section{Computation of $\mathcal M_{ab}$}
\label{app:M}

In this appendix, we compute the metric tensor of the multi-defect manifold $\mathcal M_{k\ell}$ by computing the ``overlaps". We can express the $2n \times 2n$ matrix $M_{k\ell}$ as a block matrix of four $n\times n$ matrices:
\beq \mathcal M_{k\ell} = 
\begin{pmatrix}
\mathcal M_{ij} & \mathcal N^\dag_{ij}\\
\mathcal N_{ij} & \mathcal M_{ij}
\end{pmatrix}\;,
\eeq
where
\begin{gather}
	\mathcal M_{ij} = 2\int d^2z\partial_i\bar Q_0 \bar\partial_j Q_0\\
	\mathcal N_{ij} = 2\int d^2z\bar\partial_i\bar Q_0 \bar \partial_j Q_0\;.
\end{gather}
Using the integrals below, it is easy to see that $M_{ij}$ and $\mathcal N_{ij}$ are both symmetric, and moreover, $\mathcal M_{ij}$ is real.

In the deep nematic limit, we can take $A=1$:

 \begin{eqnarray}
&{\cal M}_{ij} =\int d^2z { \sigma_j  \over {\bar z}-{\bar z}_j} { \sigma_i  \over z-z_i}\nonumber\\
&=2\int dz d{\bar z} { \sigma_i  \sigma_j \over {\bar z}[ z-(z_j-z_i)]} \nonumber \\
&=2\sigma_i  \sigma_j  \int_a^L {dR \over R} \oint\limits_R  { dz \over i[z-(z_j-z_i)]} \nonumber\\
&=4\pi \sigma_i  \sigma_j  \int_{|z_j-z_i|}^L {dR \over R}
\end{eqnarray}
so that
 \begin{eqnarray}
&{\cal M}_{ij} =4\pi \sigma_i  \sigma_j   \log {L \over \max(|z_j-z_i|,a)}\;.
\end{eqnarray}
Here $a = 0.8 \epsilon$, as we will see below in a more careful treatment, by accounting for the fact that near the defect core $A\neq 1$.

We can also calculate $\mathcal N_{ij}$, which is given by
\beq \mathcal N_{ij} = -2\int d^2z \frac{\sigma_i\sigma_j}{(\bar z - \bar z_i)(\bar z - \bar z_j)}\;.\eeq
 	We first note that for $i=j$, $\mathcal N_{ii}$ vanishes due to the phase integral. Thus below we assume $i\neq j$.
	
	Shifting $z \to z+ z_j$ and then rescaling $z \to z_{ij}z$, we have
	\beq \mathcal N_{ij} =  -2 \int  d^2z \frac{\sigma_i}{\bar z - \bar z_{ij}}\frac{\sigma_j}{\bar z} =  2\sigma_i\sigma_j\frac{z_{ij}}{\bar z_{ij}} \int   \frac{dz d{\bar z}}{{\bar z} (1-{\bar z})}\;.\eeq
		Splitting the region of integration to $|z|<1$, and $|z|>1$, and analytically expanding the integrand near $z=0$ (for $|z| < 1$) and $z=\infty$ (for $|z|>1$) yields
		\beq \mathcal N_{ij} = 2\pi \sigma_i\sigma_j\frac{z_{ij}}{\bar z_{ij}}\;.  \eeq
		Since we're interested in the large $L$ limit, $|\mathcal M_{ij}| \gg |\mathcal N_{ij}|$, and thus in this paper we will ignore $\mathcal N_{ij}$ and set $\mathcal N_{ij} = 0$.
		
\subsection{A more careful treatment of the core for $\mathcal M_{ii}$}

In order to compute $\mathcal M_{ii}$, we need to take into account the fact that near defect cores, $A\neq 1$ \cite{Pismen1999,chaikin2000principles}. We have
\begin{align}
\mathcal M_{ii} &= \int d^2z A^2\left(-\bar\partial \ln A_i - \frac{\sigma_i}{\bar z - \bar z_i}\right)\left(-\partial \ln A_i - \frac{\sigma_i}{z - z_i}\right) \nonumber\\
&{} \, + \int d^2z A^2\left(-\bar\partial \ln A_i + \frac{\sigma_i}{\bar z - \bar z_i}\right)\left(-\partial \ln A_i + \frac{\sigma_i}{z - z_i}\right) \nonumber\\
&= 2 \int d^2z A^2\left[|\partial\ln A_i|^2 + \frac{\sigma_i^2}{|z - z_i|^2}\right] \nonumber\\
&= \int d^2z \frac{1}{2} (A')^2 + 2\sigma_i^2 \int d^2z \frac{A^2}{|z - z_i|^2} \nonumber\\
&= 4\pi \sigma_i^2 \ln \frac{L}{a}\;,
\end{align}
where $a \approx 0.8\epsilon$ (using the approximate solution for $A$ in Eq.~\eqref{ASoln}).

\section{Computation of $U_i$}
\label{app:U_i}
We are interested in computing
\beq U_i = \int d^2z \bar \partial_i \bar Q \mathcal I_\alpha + \int d^2z \bar \partial_i Q \bar{\mathcal I}_\alpha = I_1 + I_2\;,\eeq
where
\begin{align}
	I_1 &= \int d^2z \bar \partial_i \bar Q[Q \partial^2 Q - (\partial Q)^2] -\int d^2z \bar \partial_i Q[ \bar Q \partial^2 Q + \partial Q \partial \bar Q]\\
	I_2 &= \int d^2z \bar \partial_i Q[ \bar Q \bar \partial^2 \bar Q - (\bar \partial \bar Q)^2 ] - \int d^2z \bar \partial_i \bar Q[Q \bar \partial^2 \bar Q + \bar \partial \bar Q \bar \partial Q]\;.
\end{align}
We compute $I_1$ and $I_2$ in order. 

\subsection{Computation of $I_1$}
Here we assume that $A = 1$, as is the case in the deep nematic limit. A more careful treatment can be found later in this appendix. 

We first note that
\begin{align}
Q_0 \partial^2 Q_0 - (\partial Q_0)^2 &= - Q_0^2\sum_j\frac{\sigma_j}{(z - z_j)^2}\\
\bar Q_0 \partial^2 Q_0 + \partial Q_0 \partial \bar Q_0 &= -\sum_j\frac{\sigma_j}{(z - z_j)^2}\;.
\end{align}
Then
\begin{align}
&\bar \partial_i \bar Q_0(Q_0 \partial^2 Q_0 - (\partial Q_0)^2) - \bar\partial_iQ_0(\bar Q_0 \partial^2Q_0 + \partial Q_0 \partial \bar Q_0) = \nonumber\\
&2Q_0 \frac{\sigma_i}{\bar z - \bar z_i}\sum_j \frac{\sigma_j}{(z - z_j)^2}\;.
\end{align}
Therefore,
\beq
I_1\approx 2\int d^2z Q_0 \frac{\sigma_i}{\bar z - \bar z_i}\sum_j\frac{\sigma_j}{(z - z_j)^2}\;. 
\eeq
We can write
\beq I_1 = I_\text{leading} + I_\text{sub}\;,\eeq
where
\begin{align}
	I_\text{leading} &= 2\int d^2z Q_0 \frac{\sigma_i^2}{|z - z_i|^2}\frac{1}{z - z_i}\\
	I_\text{sub} &= 2\int d^2z Q_0 \frac{\sigma_i}{\bar z - \bar z_i}\sum_{j\neq i}\frac{\sigma_j}{(z - z_j)^2}\;.
\end{align}

We explicitly compute and find that
\beq
I_\text{leading}=
4\pi \sigma_i^2 \alpha Q_i \delta_{2\sigma_i,1}\int dr \frac{1}{r^2}
= 4\pi \sigma_i^2 \frac{\alpha}{a} Q_i \delta_{2\sigma_i,1}\;,
\eeq
where
\beq Q_i =  \prod_{j\neq i}\frac{(z_i - z_j)^{\sigma_j}}{(\bar z_i - \bar z_j)^{\sigma_j}}\;.\eeq
We will see later in the subsection of this appendix that a more careful treatment again yields $a\approx 0.8 \epsilon$.

We now compute the subleading term $I_\text{sub}$:
\beq
I_\text{sub} \approx\sigma_i\sigma_j Q_{ij} \int d^2z \frac{(z - z_i)^{\sigma_i}}{(\bar z - \bar z_i)^{\sigma_i}}\ \frac{(z - z_j)^{\sigma_j}}{(\bar z - \bar z_j)^{\sigma_j}}\frac{1}{\bar z - \bar z_i}\frac{1}{(z - z_j)^2}\;,
\eeq
where
\beq Q_{ij} =  \prod_{r\neq i,j}\frac{(z_i - z_r)^{\sigma_r}}{(\bar z_i - \bar z_r)^{\sigma_r}}\;.\eeq
Shifting $z \to z + z_j$, we have
\beq I_\text{sub} \approx \sigma_i\sigma_j Q_{ij} \int d^2z \frac{(z - z_{ij})^{\sigma_i}}{(\bar z - \bar z_{ij})^{\sigma_i}}\ \frac{z^{\sigma_j}}{\bar z^{\sigma_j}}\frac{1}{\bar z - \bar z_{ij}}\frac{1}{z^2}\;. \eeq
Rescaling $z\to z_{ij} z$, we have
\beq I_\text{sub} =  \sigma_i\sigma_j q_{ij} \frac{I^{(1)}_{ij}}{\bar z_{ij}}\;, \eeq
where
\begin{align}
q_{ij} &= Q_i \hat z_{ij}^{2(\sigma_i - 1)}\\
I^{(1)}_{ij} &=  \int d^2z \frac{(z - 1)^{\sigma_i}}{(\bar z -1)^{\sigma_i}} \frac{z^{\sigma_j}}{\bar z^{\sigma_j}}\frac{1}{\bar z - 1}\frac{1}{z^2}\;.
\end{align}

We'll now outline how to compute $I^{(1)}_{ij}$. We first make the change of variables
\beq w^2 = \frac{z}{z-1}\;.\eeq
Then we split the region of integration to $|w|<1$, and $|w|>1$, and finally, we analytically expand the integrand near $w=0$ and $w=\infty$. Noting that $I^{(1)}_{ij}$ vanishes unless the powers of $w$ and $\bar w$ are equal yields
\begin{align} I^{(1)}_{ij} &= (-1)^{\delta_{\sigma_i + \sigma_j,1}}\frac{\pi}{1-\sigma_j}\;.
\end{align}
We want to remark that this expression is valid for $(\sigma_i,\sigma_j) = (\pm 1/2, \pm 1/2)$.

\subsection{Computation of $I_2$}

We first note that
\begin{align}
Q_0 \bar \partial^2 \bar Q_0 + \bar \partial \bar Q_0 \bar \partial Q_0 &= -\sum_j \frac{\sigma_j}{(\bar z - \bar z_j)^2}\\
\bar Q_0 \bar \partial^2 \bar Q_0 - (\bar \partial \bar Q_0)^2 &= -\bar Q_0^2\sum_j \frac{\sigma_j}{(\bar z - \bar z_j)^2}\;.
\end{align}
Then
\begin{align}
&-\bar\partial_i \bar Q_0[Q_0 \bar \partial^2 \bar Q_0 + \bar \partial \bar Q_0 \bar \partial Q_0 ] + \bar\partial_i Q_0[\bar Q_0 \bar \partial^2 \bar Q_0 - (\bar \partial \bar Q_0)^2] = \nonumber\\
&-2\bar Q_0\frac{\sigma_i}{\bar z - \bar z_i}\sum_j\frac{\sigma_j}{(\bar z - \bar z_j)^2}\;.
\end{align}
Therefore
\begin{align}
I_2&= -2\int d^2z \bar Q_0 \frac{\sigma_i}{\bar z - \bar z_i}\sum_j \frac{\sigma_j}{(\bar z - \bar z_j)^2}\;.
\end{align}
We now compute $I_2$:
\begin{align}
I_2&\approx-\sigma_i\sigma_j \bar Q_{ij} \int d^2z \frac{(\bar z - \bar z_i)^{\sigma_i}}{(z - z_i)^{\sigma_i}}\ \frac{(\bar z - \bar z_j)^{\sigma_j}}{(z - z_j)^{\sigma_j}}\frac{1}{\bar z - \bar z_i}\frac{1}{(\bar z - \bar z_j)^2}\;.
\end{align}
Shifting $z \to z + z_j$, we have
\beq
I_2= -\sigma_i\sigma_j \bar Q_{ij} \int d^2z \frac{(\bar z - \bar z_{ij})^{\sigma_i}}{(z - z_{ij})^{\sigma_i}}\ \frac{\bar z^{\sigma_j}}{z^{\sigma_j}}\frac{1}{\bar z - \bar z_{ij}}\frac{1}{\bar z^2}\;.
\eeq
Rescaling $z\to z_{ij} z$, we have
\beq I_2  = -\sigma_i\sigma_j \bar q_{ij} \frac{I^{(2)}_{ij}}{\bar z_{ij}}\;,
\eeq
where
\beq I^{(2)}_{ij} = \int d^2z \frac{(\bar z - 1)^{\sigma_i}}{(z - 1)^{\sigma_i}} \frac{\bar z^{\sigma_j}}{z^{\sigma_j}}\frac{1}{\bar z - 1}\frac{1}{\bar z^2}\;. \eeq

To compute $I^{(2)}_{ij}$, we use the same method that we used to compute $I^{(1)}_{ij}$. Doing so yields
\begin{align}
I^{(2)}_{ij} &=  \frac{\pi}{1-\sigma_j}\;.
\end{align}
Note that $|I^{(1)}_{ij}| = |I^{(2)}_{ij}|$, and the sign differs only when $\sigma_i + \sigma_j = 1$. 

\subsection{A more careful treatment of the cores for $I_\text{leading}$}

In order to quantify $a$ in the leading contribution, we need to take into account the deviation of $A$ away from 1 near the defect cores. Doing so yields
\begin{align}
	I_\text{leading} &= 2\int d^2z Q_0 A^2\left[\frac{\sigma_i^2}{|z - z_i|^2}\frac{1}{z - z_i}  -\frac{\sigma_i}{\bar z - \bar z_i}\partial^2\ln A_i \right.\nonumber\\
	&\quad \left. + |\partial \ln A_i|^2 \partial \ln A_i - \frac{\sigma_i}{\bar z - \bar z_i}(\partial \ln A_i)^2 \right.\nonumber\\
	&\left. - \frac{\sigma_i^2}{|z - z_i|^2} \partial \ln A_i + |\partial \ln A_i|^2\frac{\sigma_i}{z - z_i}\right]\;.
\end{align}
We first compute $I_\text{leading}$ by computing each term separately:
\begin{align}
&-2\int d^2z Q_0 A^2\frac{\sigma_i}{\bar z - \bar z_i}\partial^2\ln A_i \nonumber\\
& \qquad = -\pi\sigma_i \delta_{2\sigma_i,1}Q_i \int dr A\left[A A'' - (A')^2 - \frac{AA'}{r}\right]
\end{align}
\beq 2\int d^2z Q_0 A^2\frac{\sigma_i^2}{|z - z_i|^2} \frac{1}{z - z_i} = 4\pi \sigma_i^2 \delta_{2\sigma_i,1} Q_i\int dr \frac{A^3}{r^2} \eeq
\beq 2\int d^2z Q_0 A^2 |\partial \ln A_i|^2\partial \ln A_i = \frac{1}{2}\pi \delta_{2\sigma_i,1} Q_i\int dr r (A')^3\eeq
\beq -2\int d^2z Q_0 A^2 \frac{\sigma_i}{\bar z - \bar z_i}(\partial \ln A_i)^2 = -\pi\sigma_i \delta_{2\sigma_i,1}Q_i \int dr A (A')^2\eeq
\beq -2\int d^2z Q_0 A^2 \frac{\sigma_i^2}{|z - z_i|^2}\partial \ln A_i = -2\pi \sigma_i^2 \delta_{2\sigma_i,1}Q_i \int dr \frac{A^2A'}{r} \eeq
\beq 2\int d^2z Q_0 A^2 |\partial \ln A_i|^2 \frac{\sigma_i}{z - z_i} =  \pi\sigma_i\delta_{2\sigma_i,1}Q_i \int dr A(A')^2\;. \eeq
Combining all of these terms and computing, we find that
\begin{align}
&I_\text{leading}\approx \nonumber\\
&\quad 2\pi \sigma_i^2 \alpha Q_i \delta_{2\sigma_i,1}\int dr \left[-A^2A'' + A(A')^2 + \frac{2A^3}{r^2} + r(A')^3\right] \nonumber\\
&\quad \approx 4\pi \sigma_i^2 \frac{\alpha}{a} Q_i \delta_{2\sigma_i,1}\;,
\end{align}
where $a \approx 0.8\epsilon$ (using the approximate solution for $A$ in Eq.~\eqref{ASoln}).

\section{Alternative derivation of multi-defect dynamics equation}
\label{app:altMethod}

To describe nematic dynamics in the limit of weak activity and low defect density, we shall assume that the order parameter texture $Q(z,{\bar z}, t)$ stays close to the inertial manifold $Q_0 (z, {\bar z}|\{ z_i (t)\})$ parameterized by time-dependent defect positions:
\beq
Q(z,{\bar z}, t)=Q_0 (z, {\bar z}|\{ z_i (t)\})+\delta Q(z,{\bar z}, t)\;,
\eeq
where $\delta Q$ is locally perpendicular to the inertial manifold as defined by
\begin{align}
	\int dz d{\bar z} \ \partial_i\bar Q_0 \delta Q = 
		\int dz d{\bar z} \ {\bar \partial}_i\bar Q_0 \delta Q = 0\;. \label{orthogonality}
\end{align}
We thus rewrite the complex texture dynamics equation Eq.~\eqref{eq:complexQ} as
\begin{align}
{\dot z}_i \partial_i Q_0+{\dot {\bar z}}_i{\bar \partial}_i Q_0  &+\partial_t \delta Q
=\mathcal I \nonumber\\
&=-{\delta {\cal F}(\{Q\}) \over \delta \bar Q}
+\alpha \mathcal I_{\alpha}(Q)\;.
\end{align}
	Multiplying by $\partial_i\bar Q_0$ and integrating over space, we find that
	\beq \dot z_j\int dz d{\bar z}  \partial_i\bar Q_0 \partial_j Q_0  + \dot{\bar z}_j\int dz d{\bar z} \partial_i\bar Q_0 \bar \partial_j Q_0   = \int dz d{\bar z}  \partial_i\bar Q_0 \mathcal I\;. \eeq
	Similarly, if we multiply by $\bar\partial_i\bar Q_0$ and integrate over space, we find that
	\beq \dot z_j\int dz d{\bar z}  \bar\partial_i\bar Q_0 \partial_j Q_0  + \dot{\bar z}_j\int dz d{\bar z}  \bar\partial_i\bar Q_0 \bar \partial_j Q_0   = \int dz d{\bar z}  \bar\partial_i\bar Q_0 \mathcal I\;. \eeq
	If we use the physical fact that in our ansatz, $\dot{\bar z}_i$ is the complex conjugate of $\dot z_i$ (which means that in our time evolution $|Q_0|$ remains 1, which is the case in the deep nematic limit), we can combine these equations as follows by taking the complex conjugate of the first equation and adding it to the second, to get
	\beq \mathcal M_{ij}\dot z_j  + \mathcal N_{ij}\dot{\bar z}_j   = \int d^2z[\bar\partial_i\bar Q_0 \mathcal I + \bar \partial_i Q_0 \bar {\mathcal I}]\;, \label{defectEqn}\eeq
where
\begin{align}
	\mathcal M_{ij} &= \int d^2z[\bar\partial_i\bar Q_0 \partial_j Q_0 + \bar\partial_i Q_0 \partial_j \bar Q_0]\\
	\mathcal N_{ij} &= \int d^2z[ \bar\partial_i\bar Q_0 \bar \partial_j Q_0 + \bar\partial_i Q_0 \bar\partial_j \bar Q_0]\;.
\end{align}
Up to now, the discussion has been general. We will now work in the limit of small activity $\alpha \ll 1$ and large defect separation $\epsilon^{-1} \gg 1$. In this limit,  $\delta Q \ll Q_0$ because the multi-defect texture $Q_0 (z,{\bar z}|\{ z_i \})$ minimizes the LdG free energy to order ${\mathcal O} (\epsilon^2)$ on the punctured plane with fixed $z_i$. Thus to leading order
\beq \mathcal I(Q) \approx \mathcal I(Q_0)\;.\eeq
Now using the fact that
\beq \p{\mathcal F}{\bar z_i} = \int d^2z\bar\partial_i\bar Q_0 \frac{\delta \mathcal F}{\delta \bar Q_0} + \int d^2z\bar\partial_iQ_0 \frac{\delta \mathcal F}{\delta Q_0}\;,
\label{eq:Coul}\eeq
we find that
\begin{align}
&\int d^2z[\bar\partial_i\bar Q_0 \mathcal I + \bar \partial_i Q_0 \bar {\mathcal I}] = \nonumber\\
&=\int d^2z\bar\partial_i\bar Q_0 [-{\delta {\cal F} \over \delta {\bar Q_0}}+\alpha \mathcal I_\alpha(Q_0)]\nonumber\\
&+ \int d^2z\bar \partial_i Q_0[ -{\delta {\cal F} \over \delta { Q_0}}+\alpha \bar {\mathcal I}_\alpha(Q_0)] \nonumber\\
& = -\p{\mathcal F}{\bar z_i} + \alpha\int d^2z[\bar\partial_i\bar Q_0 \mathcal I_\alpha + \bar \partial_i Q_0 \bar {\mathcal I}_\alpha]\;.
\end{align}
It may seem that the substitution of $\delta {\cal F }(Q_0)/ \delta {\bar Q}=0+{\cal O} (\epsilon^2)$ on the RHS of Eq.~\eqref{eq:Coul} would lead to the vanishing of the RHS. However, one needs to worry about the integrand at the cores.

 We can check Eq.~\eqref{eq:Coul} by evaluating 
		\begin{align}
		&\int d^2z\bar\partial_i\bar Q_0 \frac{\delta \mathcal F}{\delta \bar Q_0}=-\int d^2z \bar\partial_i{\bar Q}_0 \left [ 4\partial {\bar \partial} Q_0+2\epsilon^{-2}(1-|Q_0|^2)Q_0 \right ]\nonumber\\
		&= \int d^2z {\sigma_i |Q_0|^2 \over \bar z- \bar z_i} \ 4\partial {\bar \partial} \theta  =  -4\sigma_i \int d^2z {\partial} \left({ |Q_0|^2 \over \bar z-\bar z_i}\right)\bar \partial  \theta \nonumber \\
		&= 4\pi \sum_j \frac{\sigma_i\sigma_j}{\bar z_i - \bar z_j}\;.
		\end{align}

Similarly, we find that
\beq \int d^2z\bar\partial_i Q_0 \frac{\delta \mathcal F}{\delta Q_0} = 4\pi \sum_j \frac{\sigma_i\sigma_j}{\bar z_i - \bar z_j}\;,\eeq
exactly recovering the Coulomb force term.\footnote{A more careful treatment including the core contribution gets rid of the infinity in the $i=j$ terms in the above expression.} 
Therefore upon substitution we arrive at the final equation for defect dynamics
\beq \mathcal M_{ij}\dot z_j  + \mathcal N_{ij}\dot{\bar z}_j   = -\p{{\mathcal F}_0}{\bar z_i} + {\cal U}_i\;, \eeq
with
\beq
{\cal U}_i = \alpha\int d^2z[\bar\partial_i\bar Q_0 \mathcal I_\alpha + \bar \partial_i Q_0 \bar {\mathcal I}_\alpha]\;.
\eeq
The same equation enters as the Fredholm solvability condition of the inhomogeneous linear equation for $\delta Q(z,{\bar z})$ (see Sec.~\ref{sec:method}).
It is perhaps not surprising that the equations of motion for $z_i(t)$ that we have obtained minimize the deviation of the dynamics on the inertial manifold $Q_0$ from the exact equation of motion Eq.~\eqref{eq:complexQ}. That is, we minimize
\begin{align}
E &= \int d^2z \left| \partial_t Q( z,{\bar z},t) - {d \over dt} Q_0( z,{\bar z}|\{ z_i (t) \})\right|^2 \nonumber\\
&\approx \int d^2z \left| \mathcal I (Q_0) - \dot z_i\partial_i Q_0 - \dot{\bar z}_i \bar\partial_i Q_0\right|^2
\end{align}
with respect to $\dot z_i$.

\section{Finite size corrections}
\label{app:finite}

In this appendix, we compute the finite size corrections to $\mathcal M_{ij}$ in a disc of radius $R$ with a constant boundary condition on $Q$ at $|z|=R$. We use the method of images by replacing
\beq \log \left [{z-z_i \over {\bar z}-{\bar z}_i}\right ] \to 
\log \left [{z-z_i \over {\bar z}-{\bar z}_i} \right ]+\log \left [ {z^{-1}R^2-{\bar z_i} \over {\bar z}^{-1}R^2-{ z}_i}\right ]\;,\eeq
since the added term is analytic for $|z|<R$. Since $\mathcal N_{ij}\sim\mathcal O(R^0)$, it suffices to compute corrections to $\mathcal M_{ij}$. In other words, we are interested in computing
\beq \mathcal M_{ij} = 2\sigma_i\sigma_j\int d^2z \frac{1}{z - z_i}\frac{1}{\bar z - \bar z_j} \left(1 - \frac{r^2}{R^2}\right)^2\;.\eeq

\beq I = \int d^2z \frac{\left(1 - \frac{r^2}{R^2}\right)^2}{(z - z_i)(\bar z - \bar z_j)}\frac{1}{\left(1 - \frac{z_i\bar z}{R^2}\right)\left(1 - \frac{\bar z_j z}{R^2}\right)}\eeq
		Using $z = rw$, where $w = e^{i\phi}$, and doing the contour integral over $w$, we get
		\begin{align}
		I &= 2\pi\frac{1}{1 - \frac{z_i \bar z_j}{R^2}}\left[\int_a^R {dr \over r} \ {\Theta ( r-|z_i|)- \Theta ( |z_j| - r) \over 1-r^{-2} z_i {\bar z}_j}\left(1 - \frac{r^2}{R^2}\right) \right. \nonumber\\
		&- \left.\frac{1}{R^2} \int r dr \frac{\left(1 - \frac{r^2}{R^2}\right)^2}{1 - \frac{r^2 z_i \bar z_j}{R^4}}\right]
		\end{align}
		
		Now doing the integral over $r$ yields
		\begin{align}
		\mathcal M_{ij} &= 2\sigma_i\sigma_j I \nonumber\\
		&= 2\pi \sigma_i \sigma_j
	 \left[\ln \left(\frac{R^2 - z_i\bar z_j}{r_{ij}^2}\right) - \frac{1 - \frac{r_i^2 + r_j^2}{R^2}}{1 - \frac{z_i \bar z_j}{R^2}}\right] \nonumber\\
	 &+ \sigma_i\sigma_j\frac{\pi  R^4 \left(\bar z_j z_i \left(3 \bar z_j z_i-2 R^2\right)-2 \left(R^2-\bar z_j z_i\right)^2 \log \left(1- \frac{\bar z_j z_i}{R^2}\right)\right)}{\bar z_j^3 z_i^3 \left(\bar z_j z_i-R^2\right)} \nonumber\\
	 &= 4\pi \sigma_i\sigma_j\left[ \ln \frac{R}{r_{ij}} - \frac{5}{12} + \frac{1}{4R^2} ( \bar z_j z_i \ln \frac{r_{ij}^2}{R^2} +  r_i^2 + r_j^2-\frac{11}{6} \bar z_j z_i)\right] \nonumber\\
	 &\quad + \mathcal O(1/R^4)
	 \end{align}
		
\bibliography{refs}

\end{document}